\documentclass[traditabstract,a4paper]{aa}
\usepackage{amsmath}
\usepackage{graphicx}
\usepackage{color}
\usepackage{natbib}
\usepackage[normalem]{ulem}

\bibpunct{(}{)}{;}{a}{}{,}

\begin{document}

\nocite{*}

\author{Edo van Uitert \inst{\ref{inst1},\ref{inst3}} \and Henk Hoekstra \inst{\ref{inst1}} \and Tim Schrabback \inst{\ref{inst2},\ref{inst3}} \and David G. Gilbank \inst{\ref{inst4}} \and Michael D. Gladders \inst{\ref{inst5}} \and H.K.C. Yee \inst{\ref{inst6}} }\institute{Leiden Observatory, Leiden University, Niels Bohrweg 2, NL-2333 CA Leiden, The Netherlands, email: vuitert@strw.leidenuniv.nl \label{inst1} \and Kavli Institute for Particle Astrophysics and Cosmology, Stanford University, 382 Via Pueblo Mall, Stanford, CA 94305-4060, USA \label{inst2} \and Argelander-Institut f\"ur Astronomie, Auf dem H\"ugel 71, 53121 Bonn, Germany \label{inst3} \and South African Astronomical Observatory, PO Box 9, Observatory 7935, South Africa \label{inst4} \and Department of Astronomy and Astrophysics, University of Chicago, 5640 S. Ellis Ave., Chicago, IL 60637, USA \label{inst5} \and Department of Astronomy and Astrophysics, University of Toronto, 50 St. George Street, Toronto, Ontario, M5S 3H4, Canada \label{inst6} }

\title{Constraints on the shapes of galaxy dark matter haloes from weak gravitational lensing}

\titlerunning{Constraints on the shapes of dark matter haloes}

\abstract {We study the shapes of galaxy dark matter haloes by measuring the anisotropy of the weak gravitational lensing signal around galaxies in the second Red-sequence Cluster Survey (RCS2). We determine the average shear anisotropy within the virial radius for three lens samples: the `all' sample, which contains all galaxies with $19<m_{r'}<21.5$, and the `red' and `blue' samples, whose lensing signals are dominated by massive low-redshift early-type and late-type galaxies, respectively. To study the environmental dependence of the lensing signal, we separate each lens sample into an isolated and clustered part and analyse them separately. We address the impact of several complications on the halo ellipticity measurement, including PSF residual systematics in the shape catalogues, multiple deflections, and the clustering of lenses. We estimate that the impact of these is small for our lens selections. Furthermore, we measure the azimuthal dependence of the distribution of physically associated galaxies around the lens samples. We find that these satellites preferentially reside near the major axis of the lenses, and constrain the angle between the major axis of the lens and the average location of the satellites to $\langle \theta \rangle=43.7^{\circ}\pm0.3^{\circ}$ for the `all' lenses, $\langle \theta \rangle=41.7^{\circ}\pm0.5^{\circ}$ for the `red' lenses and $\langle \theta \rangle=42.0^{\circ}\pm1.4^{\circ}$ for the `blue' lenses. We do not detect a significant shear anisotropy for the average `red' and `blue' lenses, although for the most elliptical `red' and `blue' galaxies it is marginally positive and negative, respectively. For the `all' sample, we find that the anisotropy of the galaxy-mass cross-correlation function $\langle f-f_{45} \rangle=0.23\pm0.12$, providing weak support for the view that the average galaxy is embedded in, and preferentially aligned with, a triaxial dark matter halo. Assuming an elliptical Navarro-Frenk-White (NFW) profile, we find that the ratio of the dark matter halo ellipticity and the galaxy ellipticity $f_h=e_h/e_g=1.50_{-1.01}^{+1.03}$, which for a mean lens ellipticity of 0.25 corresponds to a projected halo ellipticity of $e_h=0.38_{-0.25}^{+0.26}$ if the halo and the lens are perfectly aligned. For isolated galaxies of the `all' sample, the average shear anisotropy increases to $\langle f-f_{45} \rangle=0.51_{-0.25}^{+0.26}$ and $f_h=4.73_{-2.05}^{+2.17}$, whilst for clustered galaxies the signal is consistent with zero. These constraints provide lower limits on the average dark matter halo ellipticity, as scatter in the relative position angle between the galaxies and the dark matter haloes is expected to reduce the shear anisotropy by a factor $\sim$2.
}


\maketitle

\section{Introduction}  
\hspace{4mm} Over the last few decades a coherent cosmological paradigm has developed, $\Lambda$CDM, which provides a framework for the study of the formation and evolution of structure in the Universe. N-body simulations that are based on $\Lambda$CDM predict that (dark) matter haloes collapse such that their density profiles closely follow a Navarro-Frenk-White profile \citep[NFW;][]{Navarro96}, which is in excellent agreement with observations. Another fundamental prediction from simulations is that the haloes are triaxial \citep[e.g.][]{Dubinski91,Allgood06}, which appear elliptical in projection. This prediction of dark matter haloes, as well as many others concerning the evolution of their shapes \citep[e.g][]{Vera-Ciro11}, the effect of the central galaxy on the dark matter halo shape \citep[e.g.][]{Kazantzidis10,Abadi10,Machado10} and their dependence on environment \citep[e.g][]{Wang11}, remain largely untested observationally. \\ 
\indent Direct observational constraints on the halo ellipticities have proven to be difficult, mainly due to the lack of useful tracers of the gravitational potential. On small scales ($\sim$few kpc), halo ellipticity estimates have been obtained through the combination of strong lensing and stellar dynamics \citep[e.g.][]{VanDeVen10,Dutton11,Suyu11}, planetary nebulae \citep[e.g.][]{Napolitano10} and HI observations in late-type galaxies \citep[e.g.][]{Banerjee08,Obrien10}. On larger scales, the distribution of satellite galaxies around centrals has been used \citep[e.g.][]{Bailin08}, but such studies have only provided constraints for rich systems that may not be representative for the typical galaxy in the universe. \\
\indent Weak gravitational lensing does not depend on the presence of optical tracers and is capable of providing ellipticity estimates on a large range of scales (between a few kpc to a few Mpc). Therefore it is a powerful observational technique to study the ellipticity of dark matter haloes. In weak lensing the distortion of the images of faint background galaxies due to the dark matter potentials of intervening structures, the lenses, is measured. This has been used to determine halo masses \citep[e.g.][]{VanUitert11} as well as the extent of haloes. If galaxies preferentially align (or anti-align) with respect to the dark matter haloes in which they are embedded, the lensing signal becomes anisotropic. This signature can be used to constrain the ellipticity of dark matter haloes of galaxies \citep{Brainerd00,Natarajan00}. \\
\indent The core assumption in the weak-lensing-based halo ellipticity studies is that the orientation of galaxies and dark matter haloes are correlated; if they are not, the shear signal is isotropic and cannot be used to constrain the ellipticity of the haloes. The relative alignment between the baryons and the dark matter has been addressed in a large number of studies based on numerical simulations \citep[e.g.][]{VanDenBosch02,VanDenBosch03,Bailin05,Kang07,Bett10,Hahn10,Deason11}, in studies based on the distribution of satellite galaxies around centrals \citep[]{Wang08,Agustsson10} and in studies based on the ellipticity correlation function \citep{Faltenbacher09,Okumura09}. The general consensus is that although the galaxy and dark matter are aligned on average, the scatter in the differential position angle distribution is large. \citet{Bett11} examined a broad range of galaxy-halo alignment models by combining $N$-body simulations with semi-analytic galaxy formation models, and found that for most of the models under consideration, the stacked projected axis ratio becomes close to unity. Consequently, the ellipticity of dark matter haloes may be difficult to measure with weak lensing in practice. \\
\indent Knowledge of the relative alignment distribution is not only crucial for halo ellipticity studies, but also for studies of the intrinsic alignments of galaxies. Numerical simulations predict that the shapes of neighbouring dark matter haloes are correlated \citep[e.g.][]{Splinter97,Croft00,Heavens00,Lee08}. The shapes of galaxies that form inside these haloes may therefore be intrinsically aligned as well. Measuring this effect is interesting as it provides constraints on structure formation. Also, the lensing properties of the large-scale structure in the universe, known as cosmic shear, are affected by intrinsic alignments, and benefit from a careful characterization of the effect. Intrinsic alignments are studied observationally by correlating the ellipticities of galaxies as a function of separation; misalignments can significantly reduce these ellipticity correlation functions \citep[e.g.][]{Heymans04}. \\
\indent To date, only three observational weak lensing studies have detected the anisotropy of the lensing signal \citep{Hoekstra04,Mandelbaum_ell06,Parker07}. These studies have provided only tentative support for the existence of elliptical dark matter haloes, as they were limited by either their survey size and lack of colour information \citep{Hoekstra04,Parker07} or their depth \citep{Mandelbaum_ell06}. To improve on these constraints, we use the Red-sequence Cluster Survey 2 \citep[RCS2, ][]{Gilbank10}. Covering 900 square degree in the $g'r'z'$-bands, a limiting magnitude of $r'_{\rm{lim}}\sim 24.3$ and a median seeing of 0.7$''$, this survey is very well suited for lensing studies \citep[see][]{VanUitert11}. Using the colours we select massive luminous foreground galaxies at low redshifts. To investigate whether the formation histories and environment affect the average halo ellipticity of galaxies, the lenses are separated by galaxy type and environment, and the signals are studied separately. \\
\indent The structure of this paper is as follows. We describe the lensing analysis, including the data reduction of the RCS2 survey, the lens selection and the definition of the shear anisotropy estimators, in Section \ref{sec_analysis5}. We present measurements using a simple shear anisotropy estimator in Section \ref{sec_rat}, and use it to study the potential impact of PSF residual systematics in the shape catalogues. Various complications exist that might have altered the observed shear anisotropy, and in Section \ref{sec_multi} we study the impact of two of them: multiple deflection and the clustering of the lenses. The shear anisotropy measurements are shown and interpreted in Section \ref{sec_he}. We conclude in Section \ref{sec_concl5}. Throughout the paper we assume a WMAP7 cosmology \citep{Komatsu11} with $\sigma_8=0.8$, $\Omega_{\Lambda}=0.73$, $\Omega_M=0.27$, $\Omega_b=0.046$ and the dimensionless Hubble parameter $h=0.7$. The errors on the measured and derived quantities in this work generally show the 68\% confidence interval, unless explicitly stated otherwise.

\section{Lensing analysis \label{sec_analysis5}}
\hspace{4mm} For our lensing analysis we use the imaging data from the second Red-sequence Cluster Survey \citep[RCS2;][]{Gilbank10}. The RCS2 is a nearly 900 square degree imaging survey in three bands ({\it g$'$, r$'$} and {\it z$'$}) carried out with the Canada-France-Hawaii Telescope (CFHT) using the 1 square degree camera MegaCam. In this work, we use the $\sim$700 square degrees of the primary survey area. The remainder constitutes the `Wide' component of the CFHT Legacy Survey (CFHTLS) which we do not consider here. We perform the lensing analysis on the 8 minute exposures of the {\it r$'$}-band ({\it $r'_{lim}\sim$}24.3), which is best suited for lensing with a median seeing of 0.71$''$.
\subsection{Data reduction}
\hspace{4mm} The photometric calibration of the RCS2 is described in detail in \citet{Gilbank10}. The magnitudes are calibrated using the colours of the stellar locus and the overlapping Two-Micron All-Sky Survey (2MASS), and have an accuracy better than 0.03 mag in each band compared to the SDSS. The creation of the galaxy shape catalogues is described in detail in \citet{VanUitert11}. We refer readers to that paper for more detail, and present here a short summary of the most important steps. \\
\footnotetext[1]{http://www.cfht.hawaii.edu/Instruments/Elixir/}
\footnotetext[2]{http://www1.cadc-ccda.hia-iha.nrc-cnrc.gc.ca/cadc/}
\indent We retrieve the Elixir\footnotemark \ processed images from the Canadian Astronomy Data Centre (CADC) archive\footnotemark. We use the THELI pipeline \citep{Erben05,Erben09} to subtract the image backgrounds, create weight maps that we use in the object detection phase, and to identify satellite and asteroid trails. To detect the objects in the images, we use {\tt SExtractor} \citep{BertinA96}. The stars that are used to model the PSF variation across the image are selected using size-magnitude diagrams. All objects larger than 1.2 times the local size of the PSF are identified as galaxies. We measure the shapes of the galaxies with the KSB method \citep{Kaiser95,LuppinoK97,Hoekstra98}, using the implementation described by \citet{Hoekstra98,Hoekstra00}. This implementation has been tested on simulated images as part of the Shear Testing Programmes (STEP) (the `HH' method in Heymans et al. 2006 and Massey et al. 2007), and these tests have shown that it reliably measures the unconvolved shapes of galaxies for a variety of PSFs. Finally, the source ellipticities are corrected for camera shear, which originates from slight non-linearities in the camera optics. The resulting shape catalogue of the RCS2 contains the ellipticities of 2.2$\times 10^7$ galaxies. A more detailed discussion of the analysis can be found in \citet{VanUitert11}. \\

\subsection{Lenses \label{sec_lenses}}
\hspace{4mm} To study the halo ellipticity of galaxies, we measure the shear anisotropy of three lens samples. The first sample contains all galaxies with $19<m_{r'}<21.5$, and is referred to as the `all' sample. This sample consists of different types of galaxies that cover a broad range in luminosity and redshift. The shear anisotropy measurement of this sample enables us to determine whether galaxies are on average aligned with their dark matter haloes. \\
\begin{figure*}
  \resizebox{\hsize}{!}{\includegraphics[width=14cm,angle=-90]{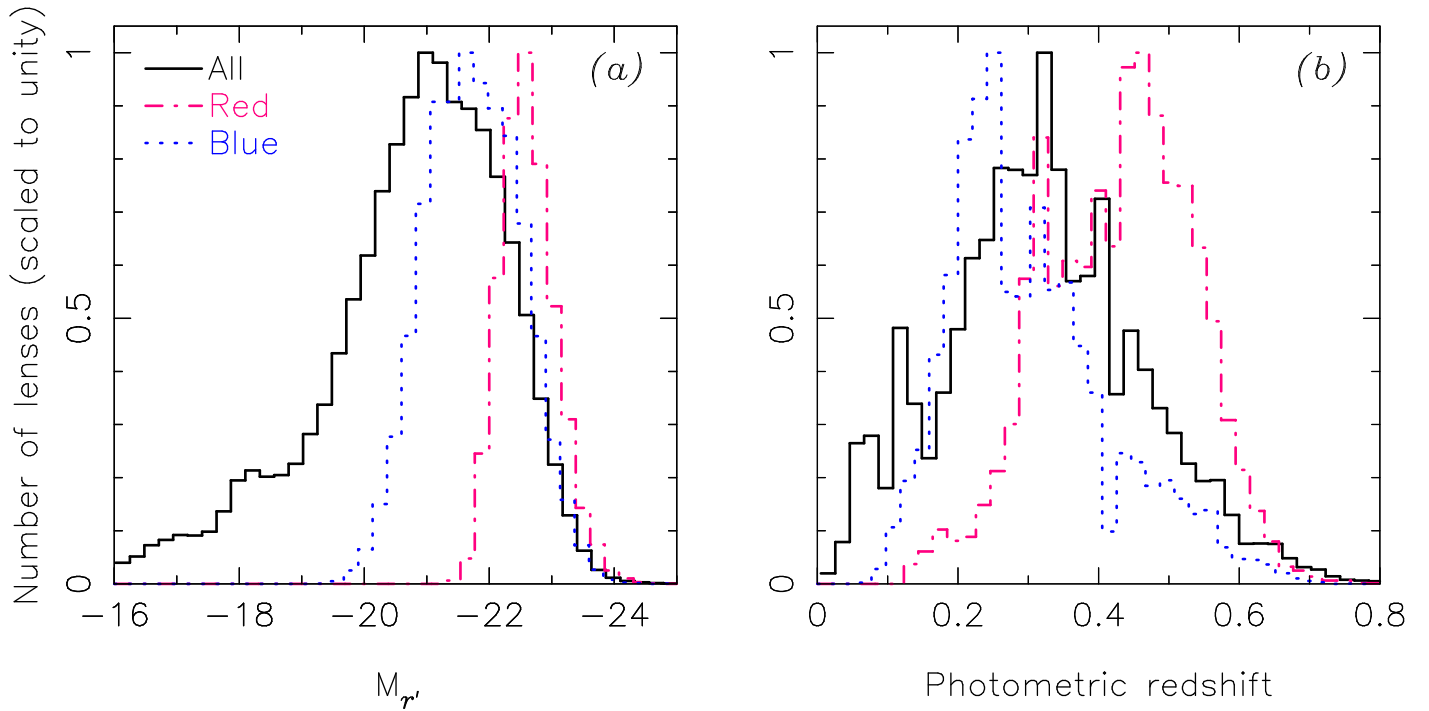}}
  \caption{Number of lenses as a function of absolute magnitude $(a)$ and redshift $(b)$ for the three lens samples, obtained by applying identical cuts to the CFHTLS W1 photometric redshift catalogue from the CFHTLenS collaboration \citep{Hildebrandt11}. The `all' sample (black solid lines) has the broadest distributions, and covers absolute $r'$-band magnitudes between $-$18 and $-$24, and redshifts between 0 and 0.6. The luminosities of the `blue' sample (blue dotted lines) are in the range $-24<M_r<-20$, with redshifts $0.15<z<0.6$. The `red' sample (purple dot-dashed lines) has the narrowest distributions, with luminosities $-24<M_r<-22$ and redshifts $0.3<z<0.6$. }
  \label{plot_galprop}
\end{figure*}
\indent The formation history of galaxies differs between galaxy types, and consequently the relation between baryons and dark matter may differ too. Therefore, the average dark matter halo shapes, and the orientation of galaxies within these haloes, might depend on galaxy type. To examine this, we separate the lenses as a function of their type. \\
\indent Various selection criteria have been employed to separate early-type from late-type galaxies. In most cases, galaxies are either selected based on the slope of their brightness profiles \citep{Mandelbaum06,VanUitert11}, or on their colours \citep{Mandelbaum_ell06,Hoekstra05}. To study how these selection criteria relate, \citet{Mandelbaum_ell06} compare the selection based on their SDSS $u-r$ model colour to the selection based on the $frac\_dev$ parameter\footnotemark, and find that the assigned galaxy types agree for 90\% of the galaxies. \\
\footnotetext[3]{The $frac\_dev$ parameter is determined by simultaneously fitting $frac\_deV$ times the best-fitting de Vaucouleur profile plus (1-$frac\_deV$) times the best-fitting exponential profile to an object's brightness profile}
\indent We choose to separate the galaxy types based on their colours, as the $g'$-, $r'$- and $z'$- band colours are readily available for all galaxies in the RCS2. The aim of the separation is two-fold: to make a clean separation between the red quiescent galaxies which typically exhibit early-type morphologies and blue star-forming galaxies that typically have late-type morphologies, and to select massive lenses at low redshifts to optimize the lensing signal-to-noise, and minimize potential contributions from multiple deflections (see Section \ref{sec_md}). To determine where these massive low-redshift galaxies reside in the colour-magnitude plane, we use the photometric redshift catalogues of the CFHTLS Wide from the CFHTLenS collaboration \citep{Hildebrandt11}, and define our boxes accordingly; details of the selection of the `red' and `blue' lens sample are described in Appendix \ref{ap_lenssel}. Note that these lens samples overlap with the `all' sample, but not with each other. Details of the samples are given in Table \ref{tab_lenssamp}. \\
\begin{table*}
  \caption{Properties of the lens samples: the number of lenses, the mean redshift, the mean luminosity, the mean ellipticity, the fraction of lenses that are isolated, the virial mass, the virial radius and the scale radius.}   
  \centering
  \renewcommand{\tabcolsep}{0.06cm}
  \begin{tabular}{c c c c c c c c c} 
  \hline
  Sample & N$_{\rm{lens}}$ & $\langle z \rangle$ & $\langle L_{r'} \rangle$ & $\langle |e_g| \rangle$ & $f_{\rm{iso}}$ & $M_{200}$ & $r_{200}$ & $r_s$ \\
   &  &  & [$10^{10}$ $h_{70}^{-2}$L$_{\odot}$] &  &  & [$10^{10}$ $h_{70}^{-1}M_{\odot}$] & [$h_{70}^{-1}$kpc] & [$h_{70}^{-1}$kpc] \\
  \hline\hline  \\
 All & 1\hspace{0.5mm}681\hspace{0.5mm}826 & 0.31 & 2.86 & 0.25 & 0.20 & $21.1^{+0.5}_{-1.4}$ & $150^{+1}_{-3}$ & $23.9^{+0.2}_{-0.5}$ \\
 Red & 136\hspace{0.5mm}196 & 0.43 & 8.91 & 0.20 & 0.41 & $138\pm8$ & $280\pm5$ & $54.6^{+1.1}_{-1.0}$ \\
 Blue & 147\hspace{0.5mm}079 & 0.31 & 4.68 & 0.26 & 0.55 & $44.4^{+3.3}_{-3.8}$ & $192\pm5$ & $32.7^{+0.8}_{-0.9}$ \\
  \hline \\
  \end{tabular}
  \label{tab_lenssamp}
\end{table*}  
\indent To study how well we can separate early-types from late-types, we compare our selection to previously employed separation criteria. Details of the comparison can be found in Appendix \ref{ap_lenssel}. We find that the `red' sample is very similar to the selection based on the $u'-r'$ colour, whilst $\sim$58\% of the `blue' sample are actually red according to their $u'-r'$ colour. Most of these contaminants of the `blue' sample are not massive, and actually dilute the lensing signal. The purity of the `blue' sample could be improved by shifting the selection boxes to bluer colours, but this at the expense of removing the majority of massive late-type lenses. Finally, we note that $\sim$70\% of the `all' sample are considered blue based on their $u'-r'$ colours. \\ 
\indent To study the second objective of the lens selection, i.e. to select massive and bright low-redshift lenses, we apply the colour cuts to the CFHTLS W1 photometric catalogue, and show the distribution of absolute magnitudes and photometric redshifts of the lens samples in Figure \ref{plot_galprop}. We find that the `red' lens sample consists of galaxies with absolute magnitudes in the range $-24<M_r<-22$, and most with redshifts between 0.3 and 0.6. The galaxies from the `blue' sample have absolute magnitudes in the range $-24<M_r<-20$, and are located at redshifts between 0.1 and 0.6. For the blue galaxies, we cannot define a criterion that exclusively selects luminous lenses in a narrow redshift range, based on the $g'r'z'$ magnitudes alone. Finally, the `all' sample has the broadest luminosity and redshift distribution. It is possible to narrow down the redshift range by discarding the lenses with the largest apparent magnitudes from each sample. We choose not to, however, because this lowers the signal-to-noise of the lensing measurement, which consequently broadens the constraints on the average halo ellipticity. \\
\indent Note that due to the lack of a very blue observing band in the CFHTLS, the photometric redshifts below 0.2 are biased high \citep{Hildebrandt11}. As a consequence, a fraction of the galaxies of the `blue' lens sample may have been shifted to higher redshifts, and thus larger luminosities. The mean redshift and luminosity of the sample may therefore be somewhat smaller than the values quoted in Table \ref{tab_lenssamp}, and the distributions shown in Figure \ref{plot_galprop} are only indicative. \\
\begin{figure}
  \resizebox{\hsize}{!}{\includegraphics[width=14cm,angle=-90]{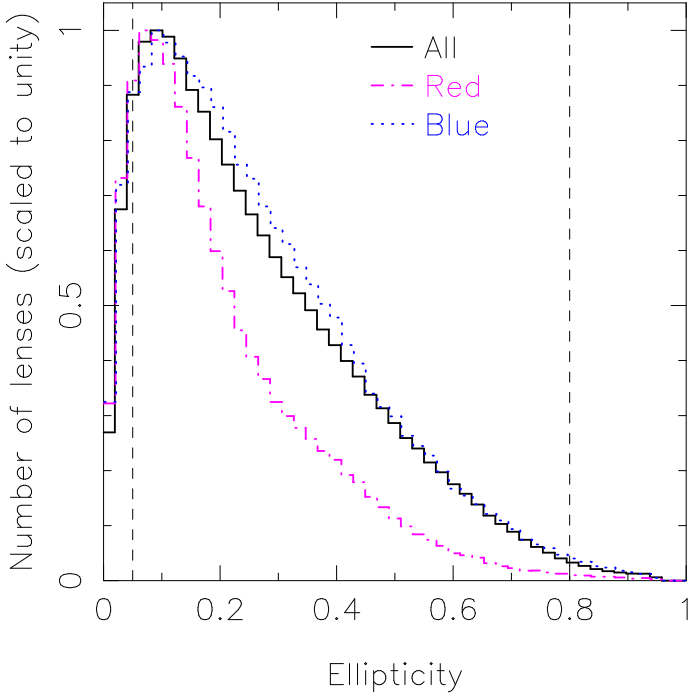}}
  \caption{Ellipticity distribution of the $g'r'z'$-colour selected lens samples. The dashed lines indicate the ellipticity cuts we apply to exclude the roundest and most elliptical lenses. The ellipticity distributions of the `all' and the `blue' sample are similar, but the `red' sample contains relatively more round galaxies.  }
  \label{plot_elldist}
\end{figure}
\indent Since the dark matter halo ellipticity is measured relative to the ellipticity of the galaxy, it is interesting to examine the  distribution of the latter. In Figure \ref{plot_elldist}, we show the ellipticity distribution of the lens samples; the mean galaxy ellipticity of each sample is given in Table \ref{tab_lenssamp}. The ellipticity distributions of the `all' and `blue' sample are comparable, and are broader than the `red' sample one, because the `all' and `blue' sample have a considerable fraction of disk galaxies. The differences between the ellipticity distributions have consequences for the weighting scheme of the lensing anisotropy measurements, as we will discuss in Section \ref{sec_shearani}. In the analysis, we only use galaxies with $0.05<e_g<0.8$, which excludes round lenses that do not have a well-defined position angle, and very elliptical galaxies whose shapes are potentially affected by neighbours and/or cosmic rays. \\
\indent The ellipticity of dark matter haloes may depend on the environment of a galaxy. We therefore divide the lens samples further into isolated and clustered ones, and study the lensing signal separately. As we lack redshifts for all the galaxies, we have to use an isolation criterion based on projected angular separations: if the lens has a neighbouring galaxy within a fixed projected separation that is brighter (in apparent magnitude) than the lens, it is selected for the clustered sample. If the lens is the brightest object, it is selected for the isolated sample. We test various values for the fixed minimum separation, and compare the tangential shear at large scales in Appendix \ref{ap_envir}. Based on these results, we use a minimum separation of 1 arcmin. Note that an environment selection based on apparent magnitudes cannot be very pure; a fraction of the lenses from the isolated sample may still be the brightest galaxy in a cluster. Some of the lenses of the clustered sample may in reality be isolated, but have been selected for the clustered sample due to the presence of bright foreground galaxies. However, the difference between the large-scale lensing signal of the isolated and the clustered sample indicates that our selection criterion works reasonably well. The fraction of the lens sample that is isolated, $f_{\rm{iso}}$, is indicated in Table \ref{tab_lenssamp}.

\subsection{Shear anisotropy \label{sec_shearani}}
\hspace{4mm}  The lensing signal is quantified by the tangential shear, $\gamma_t$, around the lenses as a function of projected separation. As the distortions are small compared to the shape noise, the tangential shear needs to be azimuthally averaged over a large number of lens-source pairs: 
\begin{equation}
  \langle\gamma_t\rangle(r) = \frac{\Delta\Sigma(r)}{\Sigma_{\mathrm{crit}}},
\end{equation}
where $\Delta\Sigma(r)=\bar{\Sigma}(<r)-\bar{\Sigma}(r)$ is the difference between the mean projected surface density enclosed by $r$ and the mean projected surface density at a radius $r$, and $\Sigma_{\mathrm{crit}}$ is the critical surface density:
\begin{equation}
  \Sigma_{\mathrm{crit}}=\frac{c^2}{4\pi G}\frac{D_s}{D_lD_{ls}},
\end{equation}
with $D_l$, $D_s$ and $D_{ls}$ the angular diameter distance to the lens, the source, and between the lens and the source respectively. Since we lack redshifts, we select galaxies with $22<m_{r'}<24$ and a reliable shape estimate as sources. We obtain the approximate source redshift distribution by applying identical magnitude cuts to the photometric redshift catalogues of the Canada-France-Hawaii-Telescope Legacy Survey (CFHTLS) ``Deep Survey" fields \citep{Ilbert06}, and find a median source redshift of $z_s=0.74$. This redshift distribution is not exactly identical to the one of the sources due to the additional shape parameter cuts applied to the source sample, which are weakly dependent on apparent magnitude, but the difference is negligble. To convert the tangential shear to $\Delta\Sigma$, we use the average critical surface density that is determined by integrating over the source redshift distribution:
\begin{equation}\begin{split}
  \langle\Sigma_{\mathrm{crit}}\rangle=\frac{c^2}{4\pi G} \frac{1}{A_{\rm{norm}}} \int_{z_l}^{\infty} dz_s \hspace{0.5mm} p(z_s)\frac{D_s}{D_lD_{ls}}; \\
 A_{\rm{norm}}= \int_{0}^{\infty} dz_s \hspace{0.5mm} p(z_s),
\end{split}\end{equation}
with $p(z_s)$ the redshift distribution of the sources, and $z_l$ the mean redshift of the lens sample used to determine $D_l$ and $D_{ls}$. We also measure the cross shear, $\gamma_{\times}$, the component of the shear in the direction of 45$^{\circ}$ from the lens-source separation vector. The azimuthally averaged cross shear signal should vanish since gravitational lensing does not produce it. If this signal is non-zero, however, it indicates the presence of systematics in the shape catalogues. As the lenses are large and their light may contaminate the lensing signal near the lenses, we only consider the signal on scales larger than 0.1 arcmin for lenses with $m_{r'}>19$, and scales larger than 0.2 arcmin for lenses with $m_{r'}<19$. These criteria are based on the reduction of the source number density near the lenses, as discussed in Appendix \ref{ap_lenslight}.  Hence the smallest scales we probe is 28 kpc for the `all' and `blue' sample, and 34 kpc for the `red' sample at the mean lens redshift. To remove contributions of systematic shear (from, e.g., the image masks), we subtract the signal computed around random points from the signal computed around the real lenses \citep[see][]{VanUitert11}. \\

\indent The lensing signal around triaxial dark matter haloes has an azimuthal dependence. If galaxies are preferentially aligned or oriented at a 90$^{\circ}$ angle (anti-aligned) with respect to the dark matter distribution, the lensing signal along the galaxies' major axis is respectively larger or smaller than along the minor axis, and this dependence can be determined. \\
\indent To measure the anisotropy in the signal, we first follow the approach used by \citet{Parker07}. For each lens, the tangential shear is measured separately using the sources that lie within $45^{\circ}$ of the semi-major axis ($\gamma_{t,\rm{B}}$), and using those that lie within $45^{\circ}$ of the semi-minor axis ($\gamma_{t,\rm{A}}$)
\begin{figure}
  \resizebox{\hsize}{!}{\includegraphics{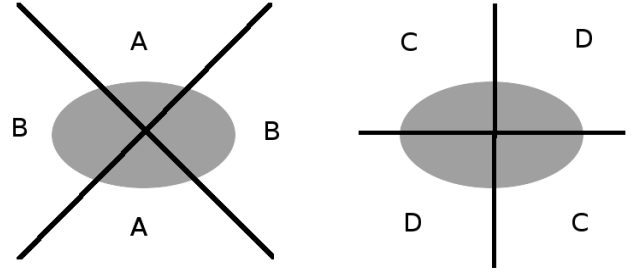}}
  \caption{Schematic of a lens galaxy. The tangential shear is measured in regions {\tt A} and {\tt B}, the cross shear is measured in regions {\tt C} and {\tt D}. The cross shear is subtracted from the tangential shear to correct for systematic contributions to the shear. }
  \label{plot_lensschematic}
\end{figure}
(indicated by {\tt B} and {\tt A} in Figure \ref{plot_lensschematic}, respectively). The ratio of the shears captures the anisotropy of the signal:
\begin{equation}
  f_{\rm{mm}}(r)=\frac{\gamma_{t,\rm{B}}(r)}{\gamma_{t,\rm{A}}(r)}.
  \label{eq_he0}
\end{equation}
A value of $f_{\rm{mm}}$ that is significantly larger (smaller) than unity at small scales indicates that the dark matter haloes are (anti-)aligned with the galaxies. Systematic contributions to the shear, however, may bias the anisotropy of the lensing signal. If the systematic shear is fairly constant on the scales where we measure the signal, it can be removed following \citet{Mandelbaum_ell06}. In this approach, the cross shear component computed in the regions that are rotated by $45^{\circ}$ with respect to the major/minor axes (region {\tt C} and {\tt D} in Figure \ref{plot_lensschematic}), $\gamma_{\times,\rm{C-D}} \equiv (\gamma_{\times,\rm{C}}-\gamma_{\times,\rm{D}})/2$, is subtracted from the tangential shear. Spurious shear signals contribute equally to $\gamma_{t,\rm{A}}$, $\gamma_{t,\rm{B}}$  and $\gamma_{\times,\rm{C-D}}$, and are therefore removed. The corrected ratio then becomes:
\begin{equation}
  f^{\rm{corr}}_{\rm{mm}}(r)=\frac{\gamma_{t,\rm{B}}(r) + \gamma_{\times,\rm{C-D}}(r)}{\gamma_{t,\rm{A}}(r) - \gamma_{\times,\rm{C-D}}(r)}.
  \label{eq_he1}
\end{equation}
If $\gamma_{\times,\rm{C-D}}$ is zero, the errors on $f^{\rm{corr}}_{\rm{mm}}$ approximately increase by a factor $\sqrt{1+1/\sqrt{2}}$; if $\gamma_{\times,\rm{C-D}}$ is non-zero, however, the errors of $f^{\rm{corr}}_{\rm{mm}}$ can either become larger or smaller than those of $f_{\rm{mm}}$. \\
\indent Alternatively, we can assume that the differential surface density distribution can be described by an isotropic part plus an azimuthally varying part \citep{Mandelbaum_ell06}:
\begin{equation}
  \Delta \Sigma_{\rm{model}} (r) = \Delta \Sigma_{\rm{iso}} (r)[1+2 f e_g \cos(2\Delta \theta)],
  \label{eq_deltasigma}
\end{equation}
where $e_g$ is the observed ellipticity of the lens, $\Delta \theta$ is the angle from the major axis, and $f$ is the ratio of the amplitude of the anisotropy of the lensing signal and the ellipticity of the galaxy, which is the parameter we want to determine.  \citet{Mandelbaum_ell06} show that the azimuthally varying part is given by:
\begin{equation}
  f \Delta \Sigma_{\rm{iso}} (r) = \frac{\sum_i w_i \Delta \Sigma_i e_{g,i} \cos(2\Delta \theta_i)}{2 \sum_i w_i e_{g,i}^2 \cos^2(2\Delta \theta_i)},
  \label{eq_f}
\end{equation}
with $i$ the index of the lens-source pairs, $w_i$ the weight applied to the ellipticity estimate of each source galaxy, which is calculated from the shape noise, and $e_{g,i}$ the ellipticity of the lens. This ellipticity is also determined using the KSB method, and it is a measure of $(1-R^2)/(1+R^2)$ with $R$ the axis ratio ($R\le1$) if the lens has elliptical isophotes. To remove contributions from systematic shear, we also measure
\begin{equation}
  f_{45} \Delta \Sigma_{\rm{iso}} (r) = \frac{\sum_i w_i \Delta \Sigma_{i,45} e_{g,i} \cos(2\Delta \theta_i+\pi/2)}{2 \sum_i w_i e_{g,i}^2 \cos^2(2\Delta \theta_i+\pi/2)},
  \label{eq_f45}
\end{equation}
where $\Sigma_{i,45}$ is the projected surface density measured by rotating the source galaxies by $45^{\circ}$. The systematic shear corrected halo ellipticity estimator is then given by  $ (f-f_{45})\Delta \Sigma_{\rm{iso}}(r)$. The average values of $f_{\rm{mm}}$, $f^{\rm{corr}}_{\rm{mm}}$ and $(f-f_{45})$ within a certain range of projected separations are determined by calculating the ratio of two measurements for each radial bin, and subsequently averaging that ratio within the range of interest. We assume that the errors of each measurement are Gaussian. Consequently, the probability distribution of the ratio is asymmetric, which we have to account for. We describe how to calculate the mean and the errors of the ratio for a radial bin, and how to average that ratio within a certain range of projected separations, in Appendix \ref{ap_asy}. Note that to convert $f$, the anisotropy in the shear field, to $f_h=e_h/e_g$, the ratio of the ellipticity of the dark matter halo and the ellipticity of the galaxy, we have to adopt a density profile (e.g. $f/f_h$=0.25 for a singular isothermal ellipsoid, see Mandelbaum et al. 2006a). \\
\indent It is clear from Figure \ref{plot_elldist} that the ellipticity distributions of the red and blue lens samples are different. It is unclear, however, whether the underlying ellipticity distribution of the dark matter haloes differs as well. If the underlying distribution is similar for both samples, the projected dark matter halo ellipticity cannot depend linearly on the galaxy ellipticity. Hence Equation (\ref{eq_deltasigma}) might not be optimal, and could depend differently on $e_g$. We therefore generalise Equation (\ref{eq_f}) to
\begin{equation}
  f \Delta \Sigma_{\rm{iso}} (r) = A \frac{\sum_i w_i \Delta \Sigma_i e^{\alpha}_{g,i} \cos(2\Delta \theta_i)}{2 \sum_i w_i e_{g,i}^{2\alpha} \cos^2(2\Delta \theta_i)},
  \label{eq_falpha}
\end{equation}
\begin{equation}
  A=\frac{ \Sigma_i e^{2\alpha}_{g,i}}{ \Sigma_i e^{\alpha}_{g,i}}\frac{ \Sigma_i e_{g,i}}{ \Sigma_i e^{2}_{g,i}}
  \label{eq_prefac}
\end{equation}
and calculate it for different values of $\alpha$. Equation (\ref{eq_f45}) changes similarly. The factor $A$ in Equation (\ref{eq_falpha}) scales each measurement of $f$ to the `standard' of $\alpha=1$ as used in \citet{Mandelbaum_ell06}, which eases a comparison of $f$ for different values of $\alpha$. The optimal weight results in the best signal-to-noise of the measurement. \\
\indent The different halo ellipticity estimators can in principle be used to study the relation between the ellipticity of the galaxy and the ellipticity of their dark matter hosts. In particular, Equation (\ref{eq_he1}) is defined such that it depends on the average dark matter halo ellipticity, whilst Equation (\ref{eq_falpha}) is sensitive to the relation between the galaxy ellipticity and the dark matter ellipticity. Hence by comparing the $f \Delta \Sigma_{\rm{iso}} (r)$ for different values of $\alpha$, we gain insight in the relation between the ellipticity of the galaxies and their dark matter haloes. Note that as an alternative, we could weight Equation (\ref{eq_he1}) with the lens ellipticity. \\
\indent It is useful to assess the signal-to-noise we expect to obtain for the shear anisotropy measurement compared to the signal-to-noise of the tangential shear itself. For this purpose, we write Equation (\ref{eq_deltasigma}) in its most basic form: 
\begin{equation}
\Delta \Sigma_{\rm{model}} (r) = \Delta \Sigma_{\rm{iso}} (r)[1+\bar{f} \cos(2\Delta \theta)], 
\label{eq_dsmodsimp}
\end{equation}
which has the following solution for the anisotropic part: 
\begin{equation}
\bar{f}\Delta  \Sigma_{\rm{iso}}=  \frac{\sum_i w_i \Delta \Sigma_{i} \cos(2\Delta \theta_i)}{\sum_i w_i \cos^{2}(2\Delta \theta_i)}.
\label{eq_fdssimp}
\end{equation}
If the dark matter halo is described by a singular isothermal ellipsoid \citep[SIE; see][]{Mandelbaum_ell06}, and if the galaxy is perfectly aligned with the halo, we find $\bar{f}=e_h/2$. Hence the anisotropic signal is a factor $e_h/2$ lower than the isotropic signal. To assess the relative size of the error of $\bar{f}\Delta \Sigma_{\rm{iso}}$ compared to $\Sigma_{\rm{iso}}$, we insert Equation (\ref{eq_dsmodsimp}) into Equation (\ref{eq_fdssimp}), define a new weight $\widetilde{w_i}\equiv w_i \cos^2(2\Delta\theta_i)$, and determine the error using $\sigma_{\bar{f}\Delta \Sigma_{\rm{iso}}}=1 / \sqrt{\sum_i \widetilde{w_i}}$. Since $w_i$ and $\cos^2(2\Delta\theta_i)$ are uncorrelated, it follows that $\sigma_{\bar{f}\Delta \Sigma_{\rm{iso}}}=1 / \sqrt{\sum_i w_i \langle \cos^2(2\Delta\theta)\rangle}=\sqrt{2}\sigma_{\Delta \Sigma_{\rm{iso}}}$, with $\sigma_{\Delta \Sigma_{\rm{iso}}}=1 / \sqrt{\sum_i w_i}$ the error on $\Delta  \Sigma_{\rm{iso}}$. Hence the error of $\bar{f}\Delta \Sigma_{\rm{iso}}$ is a factor $\sqrt{2}$ larger than the error of $\Delta  \Sigma_{\rm{iso}}$. Consequently, the signal-to-noise of the anisotropic part of the lensing signal, (S/N)$_{\rm{ani}}$, is related to the signal-to-noise of the isotropic part, (S/N)$_{\rm{iso}}$, as:
\begin{equation}
\rm{(S/N)_{ani}}=\frac{0.15}{\sqrt{2}}\left(\frac{\it{e_h}}{0.3}\right)\rm{(S/N)_{iso}}.
\end{equation}
In the best-case scenario, the expected signal-to-noise of the shear anisotropy is an order of magnitude lower than the signal-to-noise of the azimuthally averaged shear. Applying the correction to remove systematic contributions increases the errors of the shear anisotropy by another factor of $\sqrt{2}$. If the dark matter is described by an elliptical NFW, the signal decreases rapidly with increasing separation (see Figure 2 of Mandelbaum et al. 2006a), and is only larger than the SIE signal on very small scales. If no redshift information is available for the lenses, the rapid decline of the shear anisotropy is particularly disadvantageous as the signal can only be averaged as a function of angular separation. Consequently, the anisotropy signal is smeared out, making it harder to detect. Finally, if the galaxy and the halo are misaligned, the signal decreases even further. These considerations show that we need very large lens samples to achieve sufficient signal-to-noise to enable a detection, and it motivates our choice to select broad lens samples.

\subsection{Contamination correction \label{sec_contam}}
\begin{figure}
  \resizebox{\hsize}{!}{\includegraphics{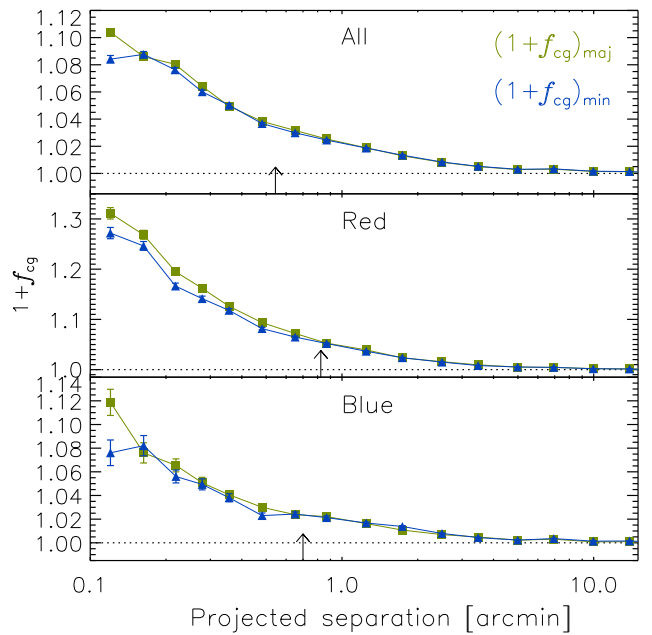}}
  \caption{Excess source galaxy density ratio as a function of projected distance to the lenses. The green squares (blue triangles) indicate the excess density ratio measured using sources within 45 degrees of the major (minor) axis. The arrows indicate the location of the virial radius at the mean redshift of the lenses. We find that the excess density ratio along the major axis is higher than along the minor axis, most noticeably for the `red' sample. Please note the different scales of the vertical axes.}
  \label{plot_od}
\end{figure}

\hspace{4mm} A fraction of our source galaxies are physically associated with the lenses. They cannot be removed from the source sample because we lack redshifts. Since these galaxies are not lensed, but are included in calculating the average lensing signal, they dilute the signal. To correct for this dilution, we boost the lensing signal with a boost factor, i.e. the excess source galaxy density ratio around the lenses, $1+f_{cg}(r)$. This is the ratio of the local total (satellites + source galaxies) number density and the average source galaxy number density. This correction assumes that the satellite galaxies are randomly oriented. If the satellites are preferentially radially aligned to the lens, the contamination correction for the azimuthally averaged tangential shear will be too low. If the radial alignment of the physically associated galaxies has an azimuthal dependence, the shear anisotropy can either be biased high or low. \\
\indent This type of intrinsic alignment has been studied with seemingly different results; some authors \citep[e.g.][]{Agustsson06,Faltenbacher07} who determined the galaxy orientation using the isophotal position angles, have observed a stronger alignment than others \citep[e.g.][]{Hirata04,Mandelbaum05} who used galaxy moments. This discrepancy was attributed by \citet{Siverd09} and \citet{Hao11} to the different definitions of the position angle of a galaxy; the favoured explanation is that light from the central galaxy contaminates the light from the satellites, which affects the isophotal position angle more than the galaxy moments one. As we measure the shapes of source galaxies using galaxy moments, we expect that intrinsic alignment has a minor impact at most and can be ignored. \\
\indent To study whether the distribution of source galaxies has an azimuthal dependence, we perform the analysis separately using the galaxies residing within 45 degrees of the major axis, and within 45 degrees of the minor axis.  On small scales, the extended light of bright lenses leads to erroneous sky background estimates, which causes a local deficiency in the source number density. This deficiency is different along the major axis and minor axis, which could bias the correction we make to account for physically associated galaxies in the source sample. To determine which scales are affected, we study the source number density around galaxies as a function of their brightness and ellipticity. The results are shown in Appendix \ref{ap_lenslight}. For galaxies with $m_{r'}<19$, we find a larger deficiency along the major axis on projected scales smaller than 0.2 arcmin; for galaxies with $m_{r'}>19$, the deficiency is larger on projected scales smaller than 0.1 arcmin. Therefore, we only use scales larger than 0.1 arcmin for galaxies with $m_{r'}>19$, and scales larger than 0.2 arcmin for galaxies with $m_{r'}<19$. The overdensities around the lens samples are shown in Figure \ref{plot_od}.
We find that the source sample is only mildly contaminated by physically associated galaxies, as the overdensities reach a maximum excess of only 30\% for the `red' lenses at the smallest projected separations. The excess source galaxy density ratio is a few percent larger along the major axis than along the minor axis within the virial radii of the lens samples, most noticeably for the `red' lens sample. \\
\begin{figure}
  \resizebox{\hsize}{!}{\includegraphics{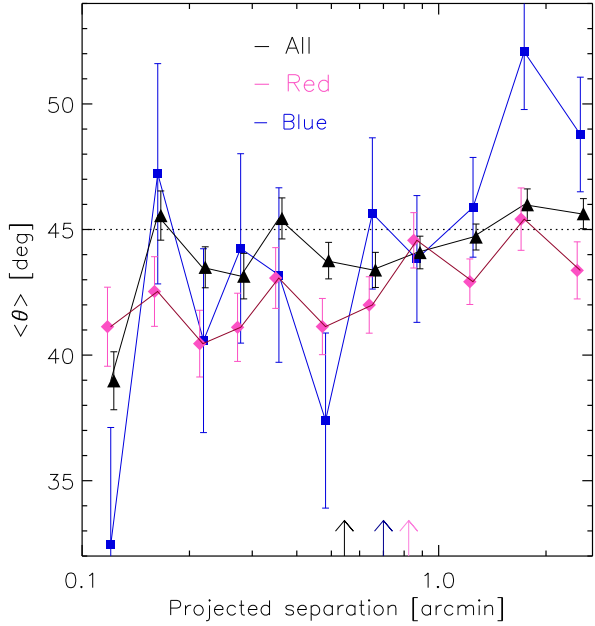}}
  \caption{Mean angle between the location of the satellites and the major axis of the lens galaxy as a function of projected separation. The black triangles, purple diamonds and blue squares indicate the results for the `all', `red' and `blue' lens sample. The arrows on the horizontal axis indicate the location of the virial radii at the mean redshift of the lenses, and correspond to 150 kpc, 280 kpc and 192 kpc for the `all', `red' and `blue' lens samples, respectively. The satellite galaxies preferentially reside near the major axis of the lenses.   }
  \label{plot_odani}
\end{figure}
\footnotetext[4]{Another effect is mentioned in \citet{Mandelbaum_ell06} that could cause an anisotropic source density ratio: additional lensing by foreground galaxies. We estimate that this has a negligible impact because the number of additional massive foreground galaxies is small due to our lens sample selection.}
\indent The measured anisotropy is caused by two effects\footnotemark: anisotropic magnification, and the presence of physically associated sources that are anisotropically distributed. As we lack redshifts for our galaxies, we cannot disentangle the two effects. However, we estimate the impact of anisotropic magnification for the lens samples in Appendix \ref{ap_magni}, and find that even in the case where the galaxy and the dark matter halo are perfectly aligned, the effect is small. We conclude therefore that the observed anisotropy is the result of the anisotropy of the distribution of satellite galaxies. \\
\indent We correct the tangential shear in the major and minor axis quadrant for the contamination by satellites by multiplying with their respective excess galaxy density ratio, before we measure the shear ratios. To calculate the correction of $(f-f_{45})$, we observe how $f\Delta\Sigma_{\rm{iso}}(r)$ changes in the presence of physically associated galaxies in the source sample that are anisotropically distributed. Rather than Equation (\ref{eq_f}), the quantity we actually measure is
\begin{equation}
  \begin{split}
  \widetilde{f}\widetilde{\Delta \Sigma_{\rm{iso}}} (r) = \frac{A}{N} \sum_i \frac{w_i \Delta \Sigma_i e^{\alpha}_{g,i} \cos(2\Delta \theta_i)}{1+f_{\rm{cg}}(r,\Delta\theta)}; \\
  N=2 \sum_i w_i e_{g,i}^{2\alpha} \cos^2(2\Delta \theta_i),
  \end{split}
  \label{eq_equation}
\end{equation}
with $1+f_{\rm{cg}}(r,\Delta\theta)$ the azimuthally varying excess galaxy density ratio, and $\widetilde{\Delta \Sigma_{\rm{iso}}}$ the unboosted lensing signal. We assume that $1+f_{\rm{cg}}(r,\Delta\theta)$ has a similar azimuthal dependence as the shear, and can be described by 
\begin{equation}
1+f_{\rm{cg}}(r,\Delta\theta)=N_{\rm{iso}}(r)+2 N_{\Delta \theta}(r) e_g^{\alpha} \cos(2\Delta\theta),
\label{eq_fcg}
\end{equation}
with $\alpha$ the exponent of the ellipticity used to weigh the shear measurement, $N_{\rm{iso}}$ the azimuthally averaged boost factor and $N_{\Delta \theta}$ the amplitude of the anisotropy. Using a Taylor expansion, we find that to first order
\begin{equation}
 f(r) =\widetilde{f(r)}+f_{\rm{eff}}(r),
\end{equation}
with $f_{\rm{eff}}(r)=A N_{\Delta \theta}(r)/N_{\rm{iso}}(r)$. To determine $f_{\rm{eff}}(r)$, we measure both the angle-averaged boost factor, $N_{\rm{iso}}(r)=N_{\rm{LS}}/N_{\rm{LR}}$, where $N_{\rm{LS}}$ denotes the number of lens-source pairs and $N_{\rm{LR}}$ the number of pairs of lenses with random sources, and the azimuthally varying part, $\xi_{\Delta \theta}(r)=\sum_{\rm{LS}} e_g^{\alpha} \cos(2\Delta \theta) / N_{\rm{LR}}$. For the adopted model of the excess galaxy density ratio this gives $N_{\rm{iso}}(r)=\langle1+ f_{\rm{cg}}(r) \rangle_{\Delta\theta}$, which is averaged over the angle, and $\xi_{\Delta \theta}=2 N_{\Delta \theta}(r) e_g^{2\alpha}$. These measurements are combined to give
\begin{equation}
f_{\rm{eff}}(r)=A\frac{\xi_{\Delta \theta}(r)}{\langle 1+ f_{\rm{cg}}(r) \rangle_{\Delta\theta}\langle e_g^{2\alpha} \rangle}.
\end{equation}
We determine the average value of $f_{\rm{eff}}(r)$ within the virial radius, and add it to $\langle f-f_{45} \rangle$. The values are tabulated in Table \ref{tab_fh}. Note that a similar correction is applied in \citet{Mandelbaum_ell06}. \\
\indent To compare the anisotropy of the distribution of satellites to the literature, we now assume that at a narrow radial range the excess galaxy density ratio can be described by $1+f_{\rm{cg}}=N_{\rm{iso}}+\widetilde{N_{\Delta \theta}} \cos(2\Delta\theta)$. We fit this to the excess density ratio in the major and minor axis quadrants, separately for each radial bin. We use these fits to compute $\langle \theta \rangle$, the mean angle between the location of the satellites and the major axis of the central galaxy, using 
\begin{equation}
\langle \theta \rangle = \frac{\int_0^{\pi/2} d\theta \theta f_{\rm{cg}}(\theta)}{\int_0^{\pi/2} d\theta f_{\rm{cg}}(\theta)}.
\end{equation}
In Figure \ref{plot_odani}, we show $\langle \theta \rangle$ as a function of projected separation for the three lens samples. \\
\indent We find that satellite galaxies preferentially reside near the major axis of the lenses, most strongly for the `red' lenses. We determine the weighted mean of $\langle \theta \rangle$ within the virial radius, and find $\langle \theta \rangle=43.7^{\circ}\pm0.3^{\circ}$ for the `all' sample, $\langle \theta \rangle=41.7^{\circ}\pm0.5^{\circ}$ for the `red' sample and $\langle \theta \rangle=42.0^{\circ}\pm1.4^{\circ}$  for `blue' sample. Additionally, for the `red' lenses we find that $\langle \theta \rangle$ becomes more isotropic at larger projected separations. It is useful to compare our results to previous studies, that are based on simulations \citep[e.g.][]{Sales07,Faltenbacher08,Agustsson10} and observations \citep[e.g.][]{Brainerd05,Agustsson06,Agustsson10,Faltenbacher07,Bailin08,Nierenberg11}. In these works, $\langle \theta \rangle$ is found to be in the range between 41$^\circ$ and 43$^\circ$ for red central galaxies, whilst no anisotropy is observed for blue central galaxies. We can only make a useful comparison for the `red' lens sample, as this sample is comparable to previously studied red galaxy samples (i.e. predominantly containing red early-type galaxies, the majority of them expected to be centrals based on their luminosity distribution). We find that the constraints agree well. For the `blue' and `all' sample, we cannot make a comparison to previous work as these samples contain a mixture of early-type and late-type galaxies, and a fair fraction of them is expected to be a satellite of a larger system. The constraints we obtained are still interesting, however, as similar selection criteria can be applied to simulations, and the results compared.


\subsection{Virial masses and radii}
\hspace{4mm} To determine to which projected separations the dark matter haloes of the galaxies dominate the lensing signal, we estimate the average halo size of each lens sample. For this purpose we model the azimuthally averaged tangential shear (after applying the contamination corrections) with an NFW profile, and fit for the mass. The NFW density profile is given by
\begin{equation}
  \rho(r) = \frac{\delta_c \rho_c}{(r/r_s)(1+r/r_s)^2},
\end{equation}
with $\delta_c$ the characteristic overdensity of the halo, $\rho_c$ the critical density for closure of the universe, and $r_s=r_{200}/c_{\mathrm{NFW}}$ the scale radius, with $c_{\mathrm{NFW}}$ the concentration parameter. We adopt the mass-concentration relation from \citet{Duffy08}
\begin{equation}
  c_{\mathrm{NFW}} = 5.71 \hspace{1mm} \Big( \frac{M_{200}}{2 \times 10^{12}h^{-1}M_{\odot}}\Big)^{-0.084} \hspace{1mm} (1+z)^{-0.47},
  \label{eq_mass_c}
\end{equation}
which is based on numerical simulations using the best fit parameters of the WMAP5 cosmology. $M_{200}$ is defined as the mass inside a sphere with radius $r_{200}$, the radius inside of which the density is 200 times the critical density $\rho_c$. We calculate the tangential shear profile using the analytical expressions provided by \citet{Bartelmann96} and \citet{WrightB00}. We fit the NFW profile between 50 and 500 kpc at the mean lens redshift; closer to the lens the lensing signal might be contaminated by lens light, and at larger separations neighbouring structures bias the lensing signal high. The best fit $M_{200}$, $r_{200}$ and $r_s$ are given in Table \ref{tab_lenssamp}. Note that in general, the best fit masses are lower than the mean halo mass because the shear of NFW profiles does not scale linearly with mass, and the distribution of the halo masses is not uniform \citep{Tasitsiomi04,Mandelbaum05HOD,Cacciato09,Leauthaud11,VanUitert11}. The resulting uncertainty in the actual mass is not important here as we are mainly interested in the extent of the haloes, which is affected less (an increase of 30\% in mass leads to an increase of only 10\% in size). \\

\section{Shear ratio \label{sec_rat}}
\hspace{4mm} In this section we present the measurements of the ensemble-averaged ratio of the tangential shear along the major and minor axis of the lenses. This is a basic indicator of the presence of anisotropies in the lensing signal. We note that the shear ratio is not an optimal estimator as the weight is simply a step function, and does not depend on the ellipticity of the galaxy. It enables, however, a comparison to \citet{Parker07}. Furthermore, we will use the shear ratio to examine how PSF residual systematics in the shape catalogues affect the anisotropy (Section \ref{sec_subPSF}). \\
\indent For all elliptical non-power law profiles, the shear ratio varies as a function of distance to the lens. This radial dependence differs for different dark matter density profiles \citep{Mandelbaum_ell06}. Hence to obtain constraints on the halo ellipticity of the dark matter, we have to adopt a particular density profile. To compare our results to those from \citet{Parker07}, we first assume that the density profile follows an SIE profile on small scales. In that case, the shear ratio is constant, and we determine the average and the 68\% confidence limits as detailed in Appendix \ref{ap_asy}. \\
\indent In Figure \ref{plot_majmin}, we show the average tangential shear along the major and minor axis, the average cross shear in the quadrants that are rotated by 45 degrees, and the inverse of the shear ratios $f_{\rm{mm}}$ and $f^{\rm{corr}}_{\rm{mm}}$. The tangential shear and the cross shear have been multiplied with the projected separation in arcmin, to enhance the visibility of the measurements on large scales where the signal is close to zero and the error bars are small. We show the inverse of the ratios following the definition used in \citet{Parker07}. We do not observe a clear signature for an alignment or anti-alignment between the lenses and their dark matter haloes. Furthermore, we find that on small scales ($<$1 arcmin), $f_{\rm{mm}}$ and $f^{\rm{corr}}_{\rm{mm}}$ are consistent, which suggests that the systematics present on these scales are smaller than the measurement errors. On larger scales, the difference is larger, which underlines the importance of applying the corrections to remove systematic contributions. The correction is largest for the `all' lens sample, because its lensing signal is smallest and therefore most susceptible to systematic contributions.
\begin{figure*}
    \includegraphics[width=1\linewidth]{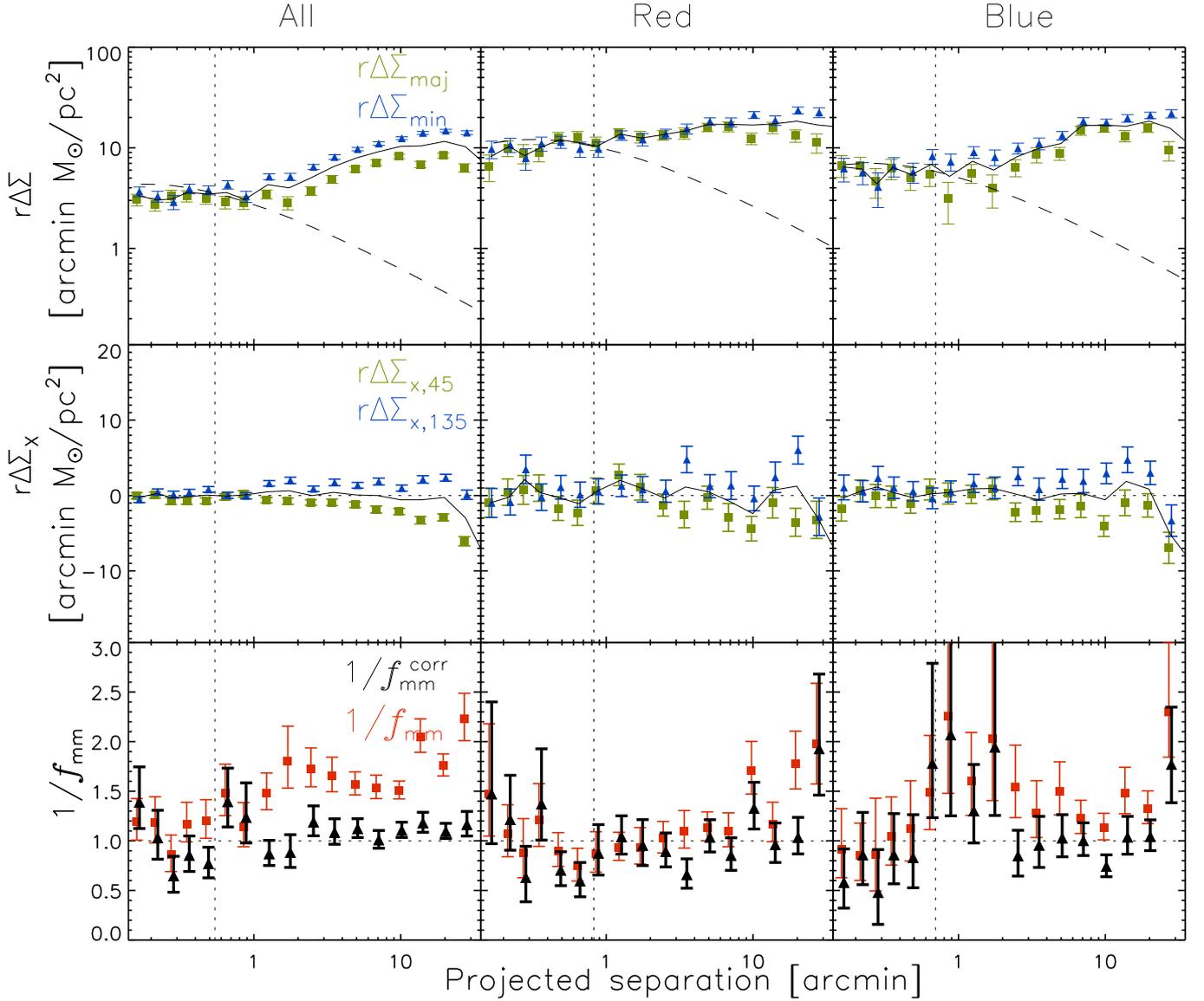}
  \caption{Lensing signal multiplied with the projected separation in arcmin as a function of angular distance from the lens, for the `all' lens sample (left-hand panels), the `red' lens sample (middle panels) and the `blue' lens sample (right-hand panels). In the top panels, the green squares (blue triangles) show the average $r\Delta \Sigma$ along the major (minor) axis (quadrants {\tt B} ({\tt A}) in Figure \ref{plot_lensschematic}). The dashed lines indicate the best fit NFW profile times the projected separation, fitted to the azimuthally averaged lensing signal on scales between 50 and 500 kpc using the mean lens redshift. In the middle panel, the green squares (blue triangles) show the cross shear signal averaged in quadrant {\tt D} ({\tt C}) of Figure \ref{plot_lensschematic}. In the bottom panels, $1/f_{\rm{mm}}$ and $1/f^{\rm{corr}}_{\rm{mm}}$ are shown by the red squares and black triangles, respectively. The dotted lines indicate the virial radius from the best-fit NFW profiles. The shear ratio does not provide clear signs for the alignment between galaxies and their dark matter haloes.}
  \label{plot_majmin}
\end{figure*}
We determine the average shear ratio within the virial radius at the mean lens redshift, and show the results in Table \ref{tab_fmm}. \\
\begin{table}
  \caption{Shear ratios for the lens samples}   
  \centering
  \begin{tabular}{c c c} 
  \hline
  Sample & $\langle 1/f_{\rm{mm}} \rangle$ & $\langle 1/f^{\rm{corr}}_{\rm{mm}} \rangle$\\
  \hline\hline  \\
 All & $1.15_{-0.09}^{+0.10}$ & $0.87\pm0.09$ \\
 Red & $0.93_{-0.09}^{+0.10}$ & $0.81_{-0.10}^{+0.11}$ \\
 Blue & $1.16_{-0.16}^{+0.19}$ & $1.04_{-0.17}^{+0.21}$\\
  \hline \\
  \end{tabular}
  \label{tab_fmm}
\end{table}     
\indent \citet{Parker07} used 22 square degrees of the CFHTLS to measure the shapes of $\sim2\times10^5$ lenses, selected with a brightness cut of $19<i'<22$. Their lens sample consisted of a mixture of early-type and late-type galaxies with a median redshift of 0.4. The shear ratio was determined using measurements out to 70 arcsec (corresponding to 250 $h^{-1}$ kpc at $z=0.4$), with a best-fit value of $\langle 1/f_{\rm{mm}}\rangle=0.76\pm0.10$. Excluding the round lenses with $e<0.15$, the best-fit ratio is $\langle 1/f_{\rm{mm}}\rangle=0.56\pm0.13$. The lens sample from \citet{Parker07} can be best compared to our `all' sample; comparing the relative number of early-/late-types in both samples using the CFHTLS W1 photometric redshift catalogue \citep{Hildebrandt11}, we find they are similar. Also, the average mass of the lenses are comparable. Fitting the shear ratio on the same physical scale, we find $\langle 1/f^{\rm{corr}}_{\rm{mm}} \rangle=0.98\pm0.08$ for the `all' sample, which is $\sim$2$\sigma$ larger than \citet{Parker07}. Excluding lenses with $e<0.15$, we find $\langle 1/f^{\rm{corr}}_{\rm{mm}} \rangle=0.95_{-0.10}^{+0.11}$, which is even almost 3$\sigma$ apart. Since the lens samples are comparable, this is most likely the result of differences in the analysis. Firstly, \citet{Parker07} do not apply a correction for systematic contributions. However, systematic shear only tends to increase $1/f_{\rm{mm}}$; if systematics were present, the discrepancy would be even larger. Secondly, it is not clear whether \citet{Parker07} accounted for the non-gaussianity of the ratio of two gaussian distributed variables in determining the shear ratio; this is particularly important when the signal-to-noise of the lensing measurements is not very high. Generally, accounting for the non-gaussianity increases the positive error bar of the shear ratio, and decreases the negative one. This could bring their result closer to ours. Finally, it is not described how the average ratio was determined. These differences could explain the discrepancy between the results.

\subsection{Imperfect PSF correction \label{sec_subPSF}}
\begin{figure}
  \resizebox{\hsize}{!}{\includegraphics[width=14cm]{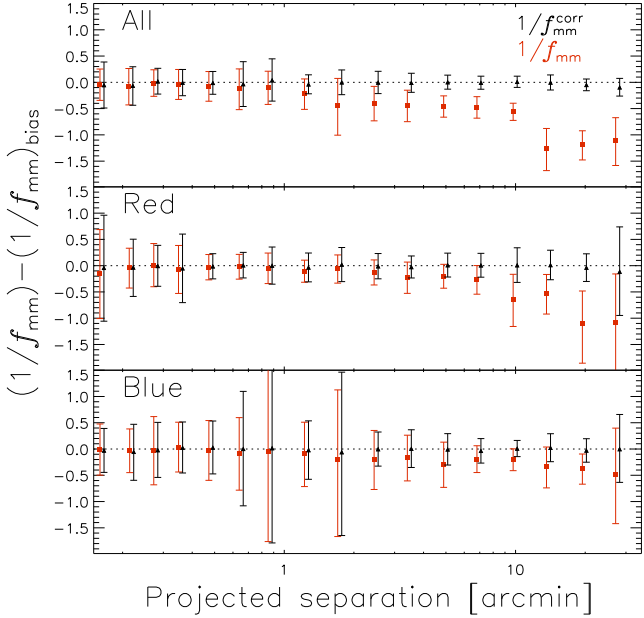}}
  \caption{Difference between the original and the PSF biased shear ratios $(1/f^{\rm{corr}}_{\rm{mm}})-(1/f^{\rm{corr}}_{\rm{mm}})_{\rm{bias}}$ (black triangles) and $(1/f_{\rm{mm}})-(1/f_{\rm{mm}})_{\rm{bias}}$ (red squares) as a function of projected separation from the lens for the three lens samples. We find that the PSF residuals are properly removed from the corrected ratio $1/f^{\rm{corr}}_{\rm{mm}}$, as the difference is consistent with zero on all scales. For the uncorrected ratio $1/f_{\rm{mm}}$, the difference is negative and decreases with projected separation. This result shows that the cross term effectively removes PSF residuals in the shear ratio estimators. }
  \label{plot_PSFtest}
\end{figure}

\hspace{4mm} To measure the ellipticities of galaxies, we have to correct their observed shapes for smearing by the PSF. The precision of the PSF correction is limited, which is mainly due to the inaccuracy of the PSF model \citep{Hoekstra04PSF}. Hence, residual PSF patterns may still be present in the shape catalogues. These residuals affect both the ellipticity estimates of the lens and the source galaxies, albeit with a different amount. Lens galaxies are typically large and bright, while source galaxies are small and faint, and hence harder to correct for. Regardless of that, PSF residuals tend to align the lens and source galaxies. If not accounted for, it could add a false anti-alignment signal to the shear anisotropy measurement (see Hoekstra et al. 2004). \\
\indent We correct for PSF residual systematics in the catalogues by subtracting the cross shear signal in the quadrants that are rotated by 45 degrees with respect to the major and minor axes ($\gamma_{x,\rm{C-D}}$ and $f_{45}\Delta_{\rm{iso}}(r)$  in $f^{\rm{corr}}_{\rm{mm}}$ and $(f-f_{45})$, respectively). To quantify how much PSF residuals actually contribute to these correction terms, and test whether they are properly removed, we introduce on purpose an additional bias in the PSF correction, and recalculate the shapes of the galaxies. Usually, the ellipticities of galaxies in the KSB method are computed as follows:
\begin{equation}
  e_g = \frac{1}{P_{\gamma}}\Big[\epsilon - (1+b)\times\frac{P^{\rm{sm}}}{P^{\rm{sm\star}}}\epsilon^{\star}\Big],
\end{equation}
with $P_{\gamma}$ the shear polarisability, $P^{\rm{sm}}$ the smear susceptibility tensor, and $\epsilon$ the polarizations \citep{Kaiser95}. The starred quantities are determined using the PSF stars. The bias $b$ is normally equal to zero, but to mimic an imperfect PSF correction we set it to $-0.05$, and recalculate the shapes of all galaxies. We create new random shear catalogues, and repeat the analysis using these biased shapes. We show the difference between the original and the biased shear ratios of the lens samples in Figure \ref{plot_PSFtest}. \\
\indent We find that the difference of the shear ratios that are determined using the original and the PSF biased catalogues is consistent with zero on all scales for $1/f_{\rm{mm}}^{\rm{corr}}$, the shear ratio estimator that is corrected with the cross shear terms. For the uncorrected shear ratio estimator, $1/f_{\rm{mm}}$, we find that the difference is consistent with zero on small scales, but turns negative for projected separations larger than a few arcmins. This shows that if PSF residuals are still present in the shape catalogues, it affects $1/f_{\rm{mm}}$, but not $1/f_{\rm{mm}}^{\rm{corr}}$. Hence we conclude that PSF residuals are properly accounted for using the cross shear signal.


\section{Impact of multiple lenses \label{sec_multi}}
\hspace{4mm} More than one lens may contribute to the shearing of a single source galaxy. Furthermore, some of the lenses are lensed themselves. In this section, we estimate the impact of these multiple lensing events on the halo ellipticity measurements. We also study the impact of the clustering of the lenses, and the correlation between their shapes, on the shear anisotropy. 

\subsection{Multiple deflections \label{sec_md}}
\begin{figure*}
  \begin{minipage}[t]{0.5\linewidth}
      \includegraphics[width=1.\linewidth]{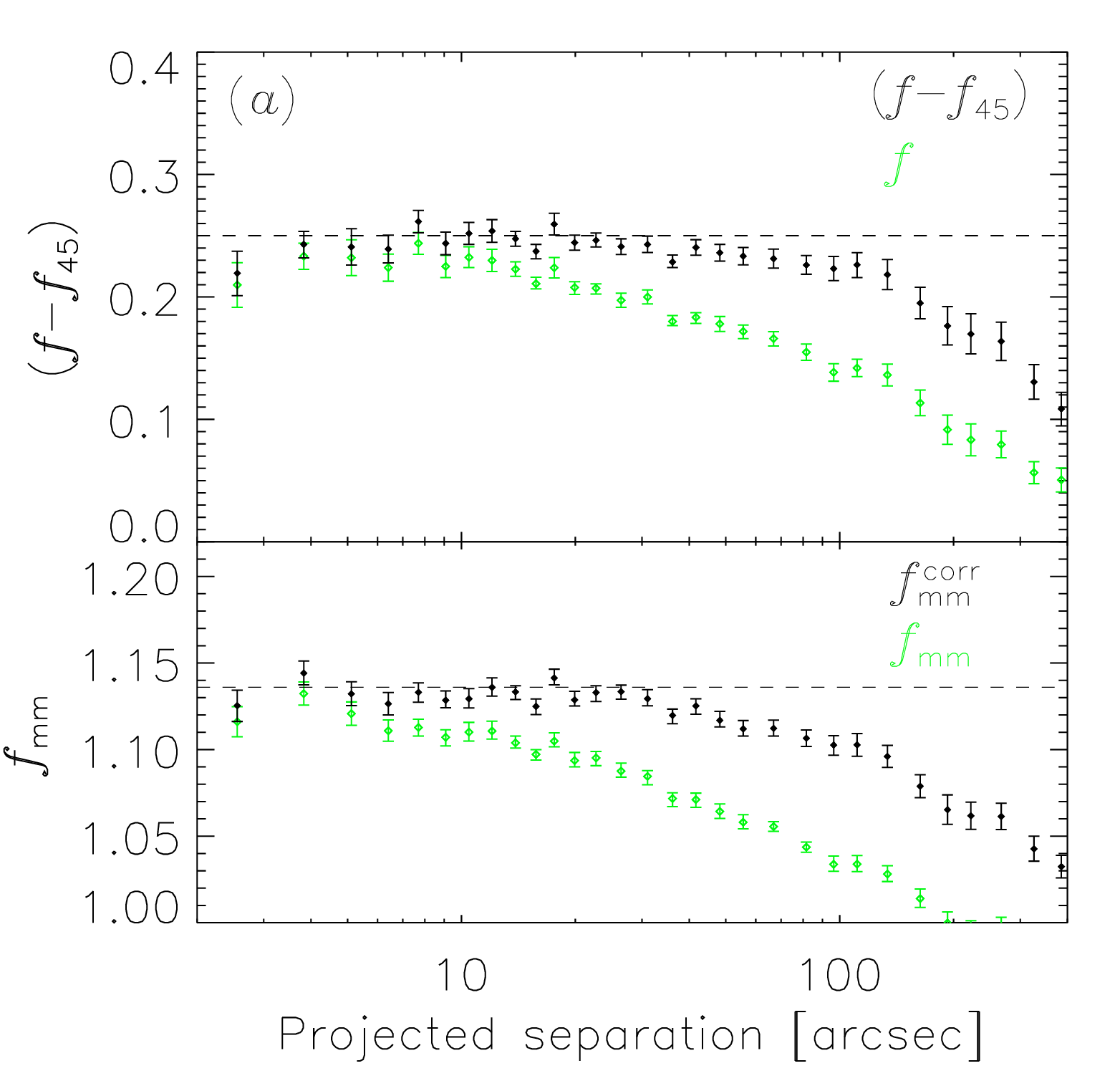}
  \end{minipage}
  \begin{minipage}[t]{0.5\linewidth}
      \includegraphics[width=1.\linewidth]{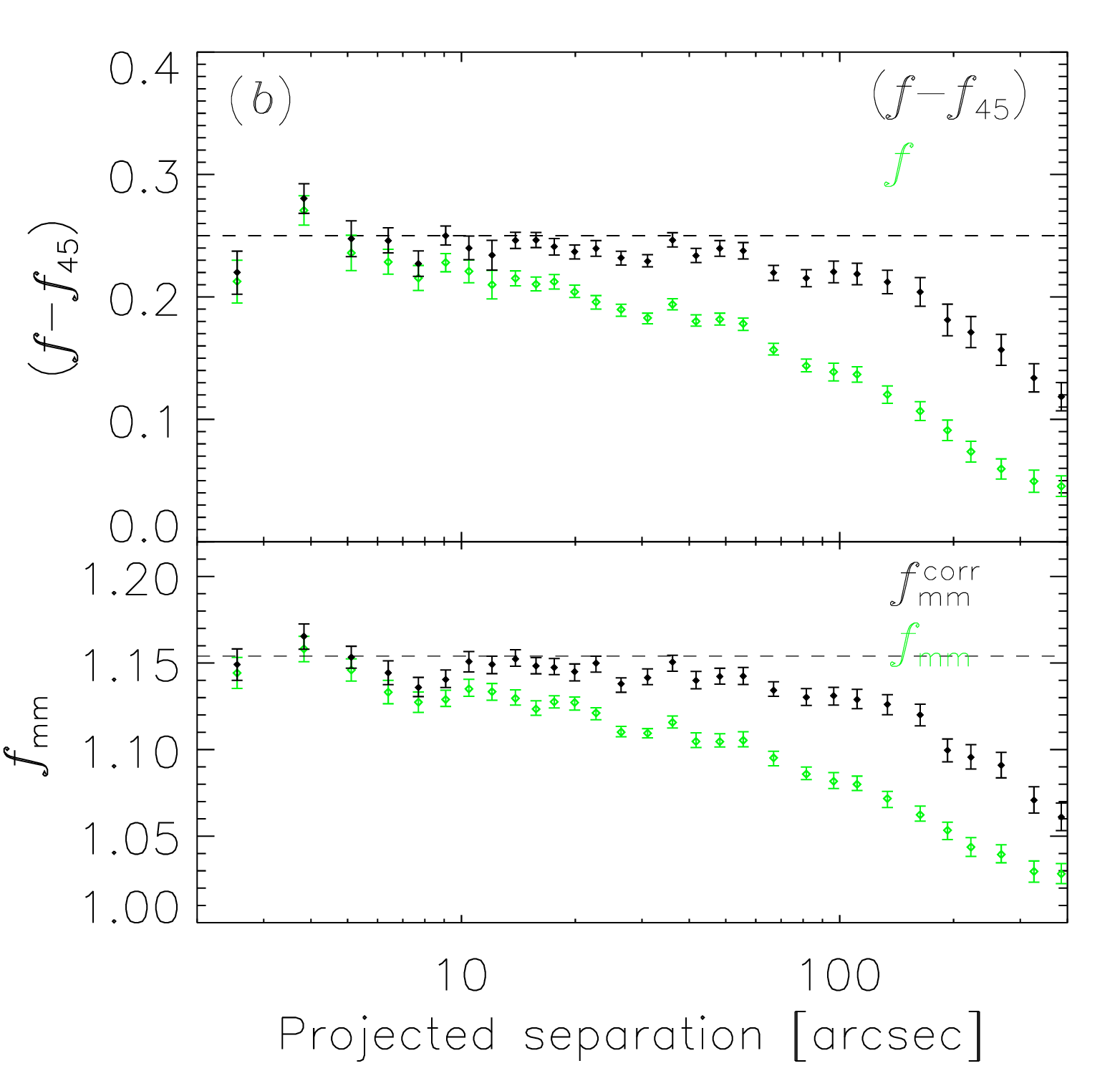}
  \end{minipage}
  \caption{$(a)$ Anisotropy of the lensing signal in the simulations in the presence of additional foreground galaxies that lens both the lenses and the sources. In the top panel, we show $(f-f_{45})$ and $f$ with the filled black and open green diamonds, computed with the exponent in the weight $\alpha=1.0$. In the lower panel, we show $f_{\rm{mm}}^{\rm{corr}}$ ($f_{\rm{mm}}$) with the filled black diamonds (open green diamonds). The dashed lines show the signal in the absence of multiple deflections. The uncorrected shear estimators $f$ and $f_{\rm{mm}}$ are increasingly reduced for larger distances to the lens. The reduction of the systematic shear corrected estimators is clearly smaller on scales $<100$ arcsec, but the difference decreases at larger separations as the systematic shear is not constant anymore and the correction therefore inaccurate. $(b)$ Anisotropy of the lensing signal in the simulations in the presence of additional foreground galaxies, excluding lens galaxies with $e<0.05$ from the analysis. The exclusion of round lenses is found to reduce the impact of multiple deflections, particularly for $f_{\rm{mm}}$ and $f_{\rm{mm}}^{\rm{corr}}$ as these measurements are not weighed with the lens ellipticity.  }
  \label{plot_modMD}
\end{figure*}

\hspace{4mm} Some foreground galaxies in our data lens both the lenses from the lens samples and the source galaxies. We denote these foreground galaxies with L2, and our selected lenses with L1. The impact of these `multiple deflections' on the halo ellipticity measurements were first discussed in \citet{Howell10}, who found that it adds a strong false anti-alignment signal to the shear anisotropy measurements. Multiple deflections affect the halo shape measurement in three ways. Firstly, the orientation of the lens light changes, leading to a misalignment if the light was aligned with the halo. Consequently, the lensing signal is averaged in quadrants that are rotated with respect to the unlensed ones, causing a reduction of the shear anisotropy. This effect is only important if the projected separation between L2 and L1 is small, as only those configurations lead to significant changes in the ellipticity of L1. Secondly, the source galaxies experience shear not only from L1, but also from L2. Especially source galaxies close to L2 are affected. Finally, the lensing of L2 changes the observed positions of L1 and the sources. We ignore this third effect as the impact is negligible. \\
\indent In the presence of L2, Equation (\ref{eq_he0}) changes to
\begin{equation}
  \tilde{f}_{\rm{mm}}(r)=\frac{\int_{-\pi/4+\delta \theta}^{\pi/4+\delta \theta}\tilde{\gamma_t}d\theta + \int_{3\pi/4+\delta \theta}^{5\pi/4+\delta \theta}\tilde{\gamma_t}d\theta}{\int_{\pi/4+\delta \theta}^{3\pi/4+\delta \theta}\tilde{\gamma_t}d\theta + \int_{5\pi/4+\delta \theta}^{7\pi/4+\delta \theta}\tilde{\gamma_t}d\theta },
\end{equation}
where the integration is performed over quadrants that are rotated by $\delta \theta$, the change of the position angle of L1 caused by the lensing of L2. $\tilde{\gamma_t}$ is the sum of the shear of L1 and L2 at the location of a source galaxy. The $\tilde{\gamma_1}$-component is given by $\tilde{\gamma_1}=\gamma_{t,L1}\cos(2\theta)+\gamma_{t,L2}\cos(2\phi)$, with $\gamma_{t,L1}$ and  $\gamma_{t,L2}$ the tangential shear of L1 and L2, $\theta$ the angle between the source galaxy and L1 and $\phi$ the angle between the source galaxy and L2. Hence the signal that is measured is given by
\begin{equation}\begin{split}
\tilde{\gamma_t}=[\gamma_{t,L1}\cos(2\theta)+\gamma_{t,L2}\cos(2\phi)]\times \cos(2\theta) + \\
[\gamma_{t,L1}\sin(2\theta)+\gamma_{t,L2}\sin(2\phi)]\times \sin(2\theta).
\label{eq_fmmtilde}
\end{split}\end{equation}
The change of the equations for $f_{\rm{mm}}^{\rm{corr}}(r)$, $f\Delta\Sigma_{\rm{iso}}(r)$ and $f_{45}\Delta\Sigma_{\rm{iso}}(r)$ in the presence of L2 can be derived in a similar way. \\
\indent To obtain an intuitive understanding of the impact of multiple deflections on the halo ellipticity measurements, we compute the change of the shear anisotropy of a single lens in the presence of an additional foreground galaxy using simple idealised simulations. These simulations, which are discussed in Appendix \ref{ap_md}, suggest that multiple deflections mainly affect the shear anisotropy of round ($e<0.15$) lens galaxies, and at large projected separations. To confirm these findings, we create a large set of simulated image catalogues to obtain a rough estimate of the impact. \\
\indent For the simulated catalogues we adopt an image size of 30$\times$30 arcmins. We randomly assign positions to 10 000 galaxies (approximately the galaxy number density of the RCS2). Redshifts are assigned to each background galaxy by drawing from the redshift distribution of the RCS2 source galaxies. The background galaxies are intrinsically round, and not convolved with a PSF, to avoid introducing unnecessary sources of noise. We insert 50 lenses at a typical lens redshift of $z=0.4$, to which we assign an ellipticity $e_{g}$ with a random value between 0 and 0.4 and a random position angle. The ellipticity of the dark matter halo, $e_h$, is proportional to the ellipticity of the galaxy via $e_h=f_{\rm{h}}\times e_g$. We use a fixed value for $f_{\rm{h}}=1.0$, but we also test the impact of loosening this assumption. Each lens is modeled with an SIE profile, and the induced shear on each source is computed with \citep{Mandelbaum_ell06}
\begin{equation}
\gamma_t=\frac{4\pi\sigma^2}{c^2}\frac{D_{l}D_{ls}}{D_s}\frac{1}{2r} \times \bigg[1+\frac{e_h}{2}\cos(2\theta)\bigg],
\label{eq_sis}
\end{equation}
with $e_h$ the ellipticity of the dark matter halo, and $\sigma$ the velocity dispersion of the lens which we have set to 200 km s$^{-1}$. We create 20 sets of 500 catalogues, and determine the mean lensing signal and the scatter between the simulations sets. This enables us to assess the significance of potential trends. \\
\indent To study the impact of multiple deflections, we would ideally assign velocity dispersions to all galaxies that reside in front of the lenses following their velocity dispersion distribution, and use them to compute the shear on the lenses and sources. The galaxies that reside behind the lenses only introduce noise, and can be ignored. This is computationally expensive as there are many galaxies at lower redshifts. However, the majority of the foreground galaxies are not massive, and are close in redshift to the lenses (resulting in small lensing efficiencies), so their contribution to multiple deflections is negligible. \\
\footnotetext[5]{Using the photometric redshift catalogue of the CFHTLS ``Deep Survey'' fields from \citet{Ilbert06}, we find that approximately 30\% of all galaxies in the RCS2 have a redshift $z<0.4$, with a mean of $z=0.35$. The lensing efficiency at $z=0.1$ is about 6 times larger than at $z=0.35$ for a source at $z=0.4$, hence we can reduce the number of L2 lenses by a factor of six when we place them at $z=0.1$. Note that we ignore the increase of the lensing efficiency by a factor of two for the source galaxies when the L2 lenses are placed at $z=0.1$, which increases the impact of multiple deflections. Additionally, the use of a smaller number of more efficient lenses also leads to an increase of the impact of multiple deflections. }
\indent For computational speed-up, we therefore define a smaller number of foreground galaxies. We choose their velocity dispersions and redshifts such that the impact of multiple deflections is comparable to what is expected using all the foreground galaxies\footnotemark. On these grounds, we randomly insert a second set of 500 round lenses with a truncated isothermal sphere (TIS) profile with a velocity dispersion of 100 km s$^{-1}$ and truncation radius of 150 arcsecs, located at a redshift of 0.1. Both the source galaxies and the L1 lenses are lensed by the L2 lenses, and we change their ellipticities accordingly. Then we measure the shear anisotropy around the L1 lenses as we would in observations, i.e. using the `observed' ellipticities. We show the anisotropy of the lensing signal in panel ($a$) of Figure \ref{plot_modMD}, confirming the predicted trends from Appendix \ref{ap_md}: multiple deflections lead to a reduction of the shear anisotropy, with a magnitude that increases for larger separations to the lens. On small scales, the reduction of the shear anisotropy is larger for the shear estimators that have not been corrected for systematic contributions ($f$ and $f_{\rm{mm}}$) than for the corrected ones ($(f-f_{45})$ and $f_{\rm{mm}}^{\rm{corr}})$ (see Figure \ref{plot_radchange}b). As long as the separation between the sources and L1 is small, the additional shear from L2 is relatively constant and hence efficiently removed using the cross terms. The correction does not work on larger scales as the additional shear from L2 varies spatially. Note that the reduction of the shear anisotropy is smaller for $(f-f_{45})$ than for $f_{\rm{mm}}^{\rm{corr}}$, because the signal of the former is weighted with the ellipticity of the lenses; the most elliptical lenses are less affected by multiple deflections (see Appendix \ref{ap_md}). \\
\indent Based on Figure \ref{plot_radchange}c in Appendix \ref{ap_md}, we expect that the impact of multiple deflections is reduced if we exclude the roundest lenses. Therefore, we repeat the simulations, excluding lenses with an observed (rather than intrinsic) ellipticity $e<0.05$ as is done in the measurements on the real data. We show the results in panel ($b$) of Figure \ref{plot_modMD}. Excluding the roundest lenses significantly reduces the impact of multiple deflections for $f_{\rm{mm}}$ and $f_{\rm{mm}}^{\rm{corr}}$. The improvement for $f$ and $(f-f_{45})$ is minor, as the roundest lenses are already downweighted in this measurement. There is some residual signal left on the largest scales, but to constrain halo shapes in real data we only use measurements on small scales. Therefore, we find it unlikely that multiple deflections strongly biases the lensing anisotropy. \\
\indent These conclusions do, however, depend on our assumptions. The number and the masses of the L2 lenses is most critical. In our simulations, we have assumed an average velocity dispersion of 100 km s$^{-1}$ for the L2 lenses. However, more massive L2 lenses contribute more to multiple deflections as their shear patterns affect larger patches of the sky. Most of the massive L2 lenses reside relatively close in redshift to the L1 lenses, and consequently their lensing efficiency is small. Therefore, we do not expect that assuming a constant velocity dispersion for the L2 lenses rather than drawing them from the distribution has a large impact on the results. \\
\indent A second important assumption is that we have modeled the density distribution of the L1 galaxy with an SIS. This is a reasonably accurate description of a galaxy density profile at small scales, but not on large scales. For NFW profiles, which are more appropriate, the lensing anisotropy declines strongly as a function of radius \citep{Mandelbaum_ell06}. A small reduction by multiple deflections of an already small signal has a relatively larger impact. One way to account for this is to give larger weights to the measurements on small scales. Alternatively, the shear anisotropy could be fit using measurements on small scales only. \\
\indent We also assumed that the relation between the ellipticity of the galaxy and the ellipticity of the dark matter halo is linear. To study the impact this might have on the anisotropy of the lensing signal, we loosen the assumption. We assign a random value to $e_h$ between 0 and 0.4, but we keep the position angles of the dark matter and the galaxy perfectly aligned. Not surprisingly, we find that the shear ratios $f_{\rm{mm}}$ and $f_{\rm{mm}}^{\rm{corr}}$ are unchanged, as the average halo ellipticity does not change. Without multiple deflections, we find that $\langle f-f_{45} \rangle$ is reduced by 25\% to 0.1875. In this case, the anisotropy of the lensing signal is no longer proportional to $e_g^2$, but to $e_h\times e_g$: the factor $e_h$ comes from Equation (\ref{eq_sis}), and the factor $e_g$ from the weight in Equation (\ref{eq_f}). When we average over a flat lens ellipticity distribution, the lensing signal in the original case is $\propto \int de_g\hspace{1mm} e^2_g$, which becomes $ \propto\langle e_h\rangle \int de_g \hspace{1mm} e_g$ if $e_h$ is randomly assigned. The second integral is 25\% smaller than the original. Hence the decrease of $\langle f-f_{45} \rangle$ results from no longer giving a larger weight to the galaxies with large dark matter halo ellipticities in the measurement. If we include multiple deflections, the relative decrease of $(f-f_{45})$ is similar as in Figure \ref{plot_modMD}a. \\
\indent Our simulations show that on small scales ($<$1 arcmin), the impact of multiple deflections is a few percent at most. This result is robust to changes in the simulation set-up, which shows that it is very unlikely that multiple deflections affect the shear anisotropy measurements on these scales. On larger scales, the impact of multiple deflection is more uncertain, and could have an important effect on weak lensing studies that investigate the alignment between galaxies and the large-scale structure. Ultimately, realistic numerical simulations should be used to accurately assess and model the impact multiple deflections have as a function of the selection of the lens sample. Such simulations are also essential to investigate the impact of large-scale structure and voids, and test whether their impact on the shear anistropy on scales close to the lens is small as expected. This is necessary to improve the precision of the gravitational lensing constraints on the shapes of dark matter haloes with data from upcoming surveys. \\
\indent \citet{Howell10} report a significant and strong decrease in $f_{\rm{mm}}$ with projected lens-source separation. However, for computational speed-up, \citet{Howell10} only use those foreground galaxies that reside within 100 arcsec from the lens to calculate the multiple deflection signal. Figure \ref{plot_radchange}a shows that this leads to an overestimation of the reduction of the shear ratios. Furthermore, redshifts are assigned to galaxies whose positions have been observed in real data. This causes an unphysical correlation between the positions of foreground and background galaxies, and amplifies the chance of having a low-redshift foreground galaxy close to a lens, which could result in an overestimate of the effect. Finally, no correction for systematic shear is implemented, which would have reduced the impact as well. \\

\subsection{Clustering of lenses} 
\begin{figure*}
  \begin{minipage}[t]{0.5\linewidth}
      \includegraphics[width=1.\linewidth]{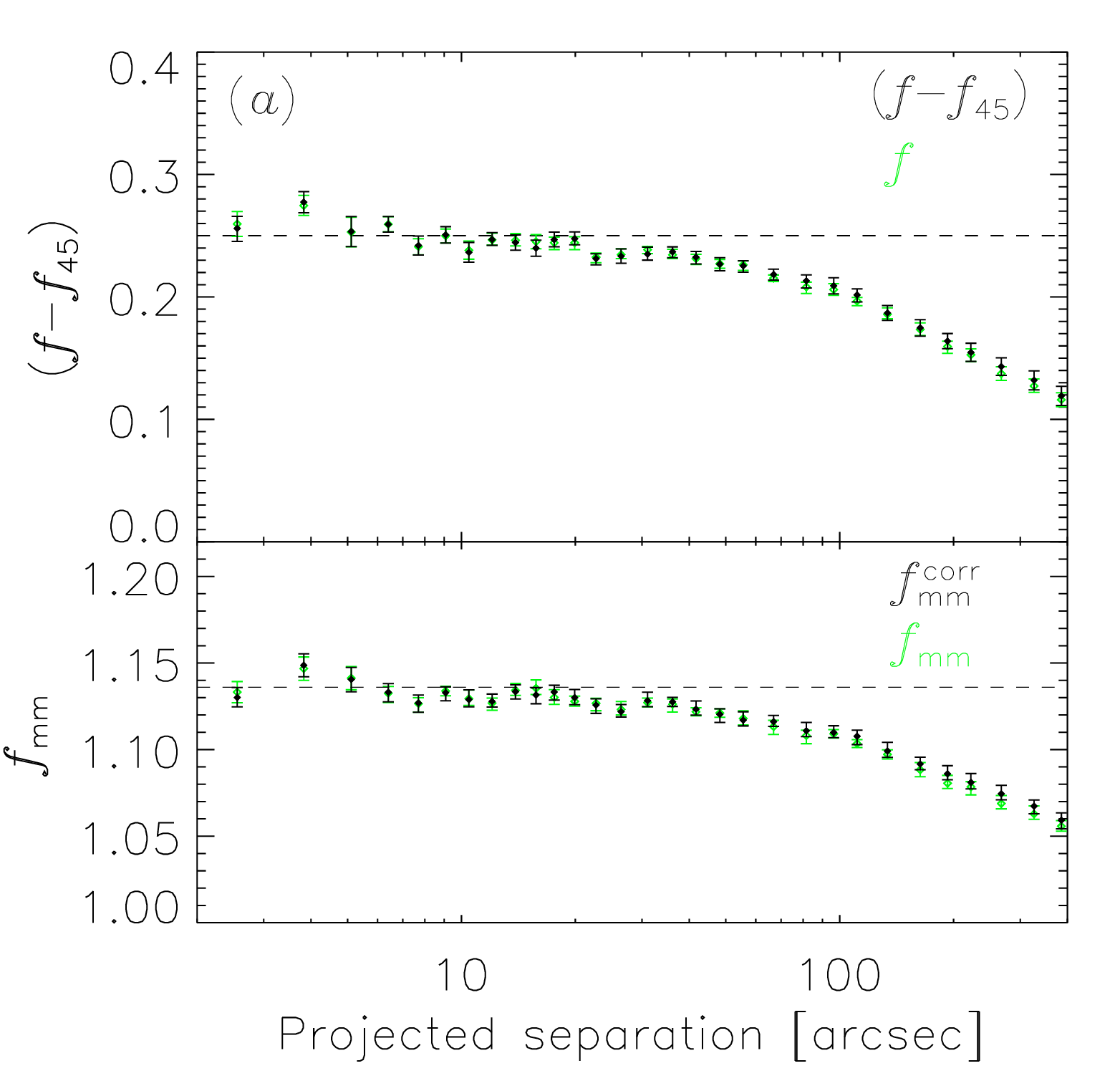}
  \end{minipage}
  \begin{minipage}[t]{0.5\linewidth}
      \includegraphics[width=1.\linewidth]{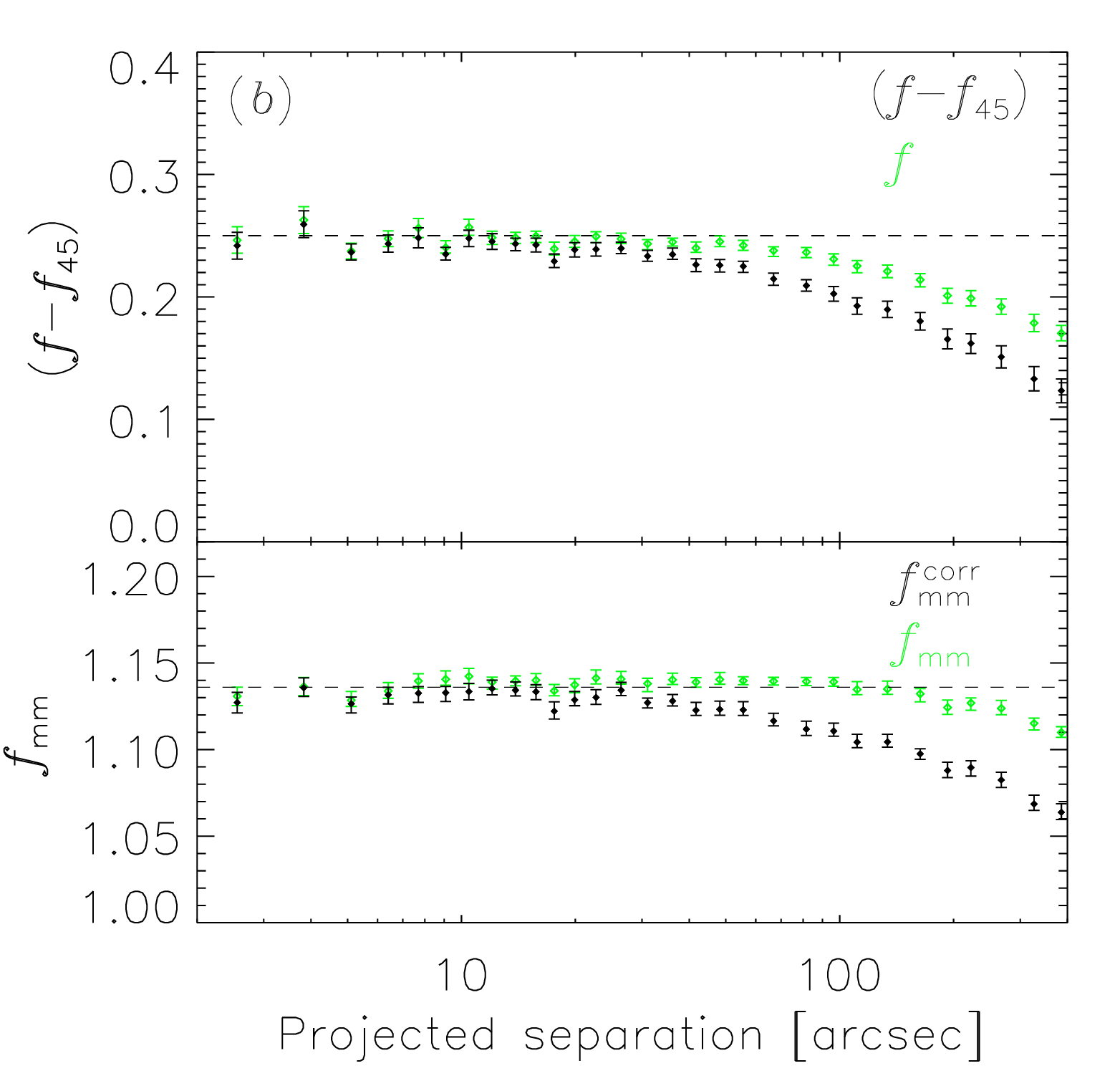}
  \end{minipage}
  \caption{($a$) Anisotropy of the lensing signal in the simulations using the observed positions of the `red' lens sample in the RCS2. The filled black diamonds (open green diamonds) show the shear anisotropy estimators (not) corrected for systematic shear. The ellipticities of the lenses have been randomly assigned. The lensing anisotropy declines slightly at scales $>20$ arcsec, which is expected as clustering of lenses with random position angles leads to an isotropic signal at large scales. In panel ($b$), we add an additional ellipticity of 0.03 to the $e_1$ component of the lenses to mimic intrinsic alignments. This increases the shear anisotropy, but also induces cross shear, so that the systematic shear corrected shear anisotropy estimators are unchanged. }
  \label{plot_clus}
\end{figure*}
\hspace{4mm} Many galaxies in the Universe reside in groups and clusters. Hence the shear we measure around a lens galaxy is the sum of the shear from the lens and neighbouring galaxies. At small projected separations from the lens, the signal is dominated by the lens galaxy, but at larger separations the shear from the neighbours becomes increasingly important. This may affect the shear anisotropy measurements. To study this, we assign the positions of the lenses from the real lens samples to the simulated lenses. The lens ellipticities are randomly drawn between 0 and 0.3, and are modeled with an SIE profile. The positions of the galaxies with $22<m_{r'}<24$ in the RCS2 are assigned to the simulated sources. The lenses are put at a redshift $z=0.4$, and the source redshifts are randomly drawn from their redshift distribution. We assume that $f_{\rm{h}}=1$, and that no foreground lenses are present as that would potentially mix different effects. To obtain the errors, we determine the scatter between 20 random realizations. We show the results in Figure \ref{plot_clus}a for the `red' lens sample.  \\
\indent We find that clustering of the red lenses slightly reduces the lensing anisotropy on large scales. This can be easily understood: if galaxies with random ellipticities cluster, the resulting lensing signal around the lens ensemble at large scales becomes more isotropic. The impact of clustering on the shear anisotropy around the `blue' and `all' lenses is smaller, and is therefore not shown. We note that, additionally, dark matter haloes may be stripped in high-density environments, which could reduce the anisotropy signal as well. \\
\indent Numerical simulations suggest that dark matter haloes are aligned, and point toward each other \citep[e.g.][]{Splinter97,Croft00,Heavens00,Lee08}. If the galaxies that reside in these haloes are preferentially aligned with the halo - a prerequisite for measuring halo shapes - the observed galaxy ellipticities are expected to show this alignment as well. In particular, the ellipticities of luminous red galaxies are believed to be increasingly correlated with decreasing separations \citep[e.g.][]{Hirata07,Okumura09,Joachimi11}, which can be understood in a framework in which galaxies form in a linear tidal field \citep[][]{Blazek11}. Similar studies for spiral galaxies show the effect is considerably weaker \citep[][]{Mandelbaum11}. Intrinsic alignments also affect the shear anisotropy signal.\\
\indent To obtain a conservative estimate of the impact of intrinsic alignments, we add an ellipticity of 0.03 to all the $e_1$ components of the lenses. The result for the `red' lenses is shown in Figure \ref{plot_clus}b. We find that correlated lens ellipticities increases the shear anisotropy on large scales; the shear pattern of neighbouring lenses amplifies each other. However, correlated lens ellipticities also induces cross shear of an equal magnitude, hence the corrected shear anisotropy estimator is unchanged. In reality, the intrinsic alignments are scale dependent, and the correlation decreases for larger separations. Furthermore, if the lens galaxies are pointing towards each other, the contributions of the shear along the major axis systematically add up, which could amplify the effect. This is difficult to model, however, but we will assess it with numerical simulations in a future work.

The main conclusion from our simulations is that the impact of multiple deflections, clustering of lenses and the correlations of their ellipticities is small, and can be safely ignored for the lens selection in this work (i.e. particularly for the `blue' and `red' lens sample, which were selected to contain massive, elliptical, and low-redshift galaxies). More realistic studies using numerical simulations are required to quantify the effects more precisely. This is a crucial step in correctly interpreting the lensing anisotropy in future lensing data that are of higher precision.

\section{Halo ellipticity \label{sec_he}}
\begin{figure*}
  \resizebox{\hsize}{!}{\includegraphics{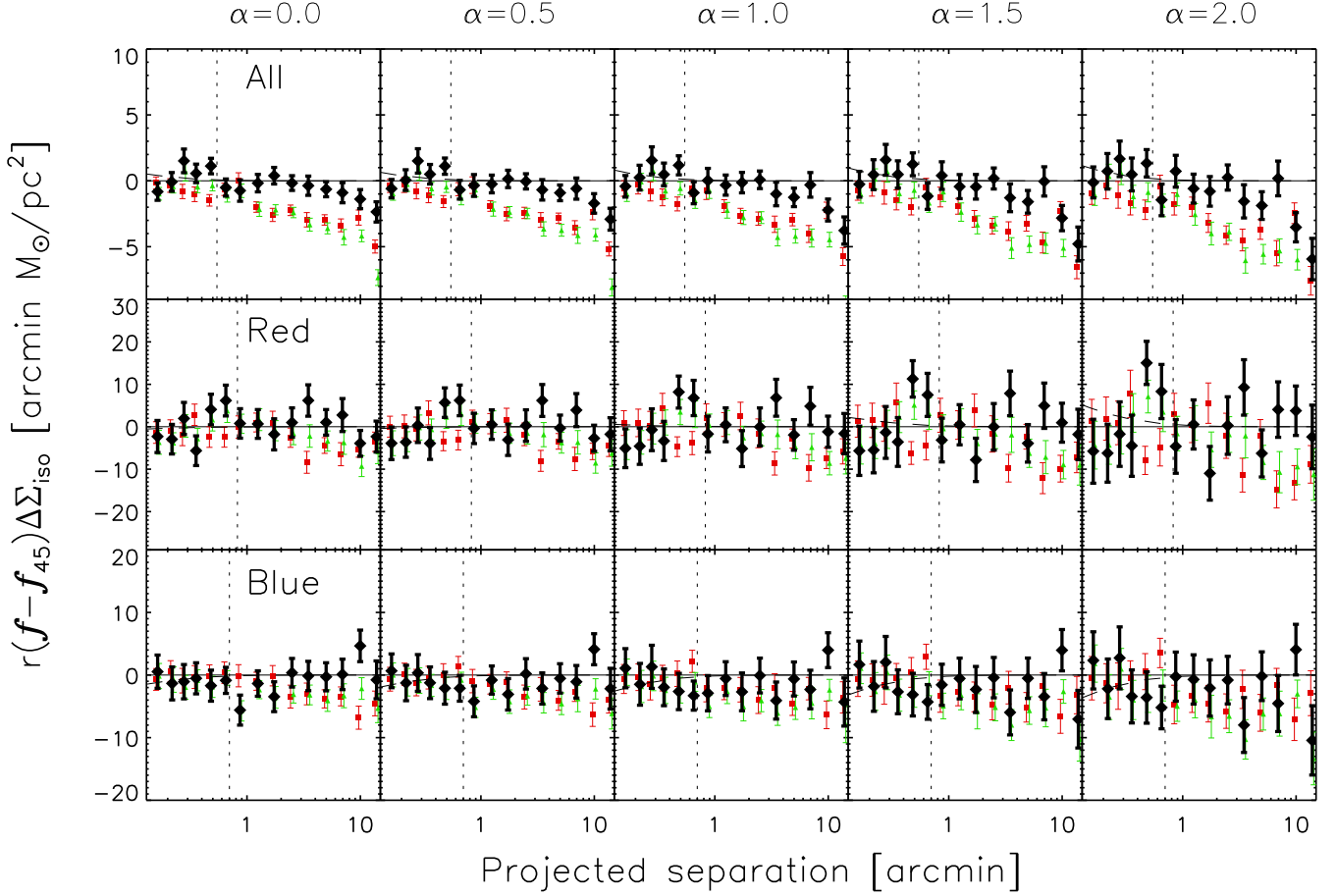}}
  \caption{ Anisotropic shear signal around the `all' lenses (top panels), the `red' lenses (middle panels) and the `blue' lenses (bottom panels). The signal has been multiplied with the projected separation in arcmin to enhance the visibility of the signal on large scales. The green triangles denote $r\hspace{0.5mm} f \Delta \Sigma_{\rm{iso}}$, the red squares $r\hspace{0.5mm}f_{45} \Delta \Sigma_{\rm{iso}}$, and the black diamonds $r(f-f_{45}) \Delta \Sigma_{\rm{iso}}$. The vertical dotted lines indicate the virial radius at the mean lens redshift, the dashed lines show the best fit elliptical NFW profiles. Indicated on top of each column is the weight applied to the measurement. The average shear anisotropy of the `all' and `red' lenses is weakly positive within the virial radius, supporting the presence of triaxial dark matter haloes that are reasonably aligned with the galaxies. For the `blue' lenses the signal is marginally negative. Note that the measurements shown here are not corrected for the anisotropic distribution of satellite galaxies. }
  \label{plot_shearani}
\end{figure*}
\hspace{4mm} The simulations from the previous section indicate that multiple deflections and clustering have a small impact on the shear anisotropy, in particular for the `blue' and `red' lens samples where we have selected massive, elliptical, low-redshift lenses. Now we proceed with the actual measurements. We show $(f-f_{45})\Delta \Sigma_{\rm{iso}}$ as a function of projected distance to the lens in Figure \ref{plot_shearani}. We multiply the lensing signal with the projected separation in arcmin to enhance the visibility of the signal on large scales. Each row shows the measurements for one of the lens samples, whilst the exponent of the galaxy ellipticities, $\alpha$ in Equation (\ref{eq_falpha}), differs between the columns. To quantify the shear anisotropy, we first determine the average value of $(f-f_{45})$ as detailed in Appendix \ref{ap_asy}. The best-fit values are summarized in Table \ref{tab_fh}. \\
\indent For the `all' sample of lenses, we find that $\langle f-f_{45} \rangle$ is marginally positive, independent of the applied weight. For the `red' lenses, we find that $\langle f-f_{45} \rangle$ is consistent with zero if we take $w\propto e^{0.0}$. For larger values of $\alpha$, however, $\langle f-f_{45} \rangle$ turns marginally positive due to the two data points near the virial radius. Closer to the lens, the signal is slightly negative. For the blue galaxies, the signal is marginally negative. \\
\indent Next, we fit an SIE and an elliptical NFW profile. For the SIE fit, we first determine the Einstein radius $r_{\rm{E}}$ by fitting a singular isothermal sphere (SIS) profile to the azimuthally averaged tangential shear measurements within the virial radius at the mean lens redshift. To determine $f_{\rm{h}}$, we fit \citep{Mandelbaum_ell06}
\begin{equation}
(f-f_{45})\Delta \Sigma(r)=\frac{\Sigma_{\rm{crit}}r_{\rm{E}} f_{\rm{h}}}{8r};
\end{equation}
the errors of $f_{\rm{h}}$ are determined from the $\chi^2$-values of the fit. Since $(f-f_{45})/f_{\rm{h}}=0.25$ on all scales for an SIE profile \citep{Mandelbaum_ell06}, we expect to find best-fit values for $f_{\rm{h}}$ that are four times larger than $\langle f-f_{45} \rangle$. The best-fit values are shown in Table \ref{tab_fh}. We find that $f_{\rm{h}}$ is consistent with, but not exactly, four times $\langle f-f_{45} \rangle$. The difference may be due to differences between the fitting methods; if we first fit an SIS profile, we correlate the azimuthally averaged tangential shear measurements, whilst if we determine $\langle f-f_{45}\rangle$ directly from the data, each data point is treated separately. The general trends, however, are consistent and the conclusions do not depend on how we fit the data. \\
\indent Currently, no analytical expression exists for the shear anisotropy of an elliptical NFW profile. Therefore, we use the numerically integrated values of $f/f_h$ and $f_{45}/f_h$ as a function of $r/r_s$ from \citet{Mandelbaum_ell06} (shown in Figure 2 of that paper), which have been kindly provided by Rachel Mandelbaum. Since our lens galaxies span a broad range in redshifts, we first determine the redshift-averaged lensing model by integrating the elliptical NFW profiles over the redshift distribution of each lens sample (shown in Figure \ref{plot_galprop}), and weigh each lens redshift bin with the lensing efficiency $\langle D_{ls}/D_{s} \rangle$ that is averaged over the source redshift distribution. Note that this is an important correction; for the `all' sample, the integrated profile results in about 50\% larger values for $(f-f_{45})$ compared to the profile computed using the mean lens redshift. For the `red' and `blue' sample, the difference is smaller because their redshift distributions are narrower. Also note that the SIE profiles do not have to be corrected, since the azimuthally averaged lensing signal and the anisotropic part are similarly affected when we integrate them over the lens redshift distribution because both scale as $r^{-1}$. The best-fit values of $f_h$ are therefore unaffected.  \\
\indent The best fit values of $f_h$ for the elliptical NFW profiles are less significant than $\langle f-f_{45} \rangle$ for the same lensing measurements. The reason is that the elliptical NFW fit is very sensitive to the signal close to the lens, but not to the signal at larger separations. We find that for the `all' and `red' sample, $(f-f_{45})$ actually turns slightly negative close to the lens, rather than increasing strongly as would have been expected for an elliptical NFW profile that is aligned with the lens. Although this might be just caused by noise, it could also indicate that a single elliptical NFW profile does not describe the shear anisotropy signal well. \\
\indent Finally, we note that for the `all' sample $\langle f-f_{45} \rangle \Delta\Sigma_{\rm{iso}}$ turns negative at projected separations $>$5 arcmin. A similar trend can be observed in Figure \ref{plot_majmin}, where the inverse of the corrected shear ratio of the `all' sample is slightly larger than unity.  We cannot directly interpret this as the result of an anti-alignment of galaxies with the large-scale structure, as we found in the previous section that multiple deflections and the clustering of galaxies produce a similar trend at these scales, and we cannot disentangle the effects.  To constrain the average halo ellipticities of galaxies we only use the lensing signal on scales $<$1 arcmin, however, where the effect of multiple deflections and clustering of galaxies can be safely ignored, and a non-zero signal reflects an anisotropy of the projected gravitational potential.
\begin{table*}
  \caption{The best-fit values for the anisotropy of the galaxy-mass cross-correlations function, $\langle f-f_{45} \rangle$, and the ratio of the dark matter halo ellipticity and the galaxy ellipticity, $f_{\rm{h}}$, for an SIE and an elliptical NFW profile.}   
  \centering
  \begin{tabular}{c c c c c c } 
  \hline
  Sample  & $\alpha$ & $\langle f_{\rm{eff}} \rangle $ & $\langle f-f_{45} \rangle$ & $f_{\rm{h}}$(SIE) & $f_{\rm{h}}$(NFW) \\
   &  &  &  &   & \\
  \hline\hline  \\
  All &  $0.0$ & $1.3\pm0.6\times10^{-3}$ & $0.19\pm0.10$ & $0.47\pm0.37$ & $0.96_{-0.80}^{+0.83}$ \\
  All &  $0.5$ & $1.1\pm0.7\times10^{-3}$ & $0.21_{-0.10}^{+0.11}$ & $0.57\pm0.40$ & $1.19_{-0.85}^{+0.89}$ \\
  All &  $1.0$ & $0.8\pm0.8\times10^{-3}$ & $0.23\pm0.12$ & $0.70\pm0.46$ & $1.50_{-1.01}^{+1.03}$ \\
  All &  $1.5$ & $0.6\pm1.0\times10^{-3}$ & $0.26\pm0.15$ & $0.83\pm0.55$ & $1.80_{-1.19}^{+1.23}$ \\
  All &  $2.0$ & $0.4\pm1.2\times10^{-3}$ & $0.29\pm0.17$ & $0.97\pm0.65$ & $2.12_{-1.42}^{+1.45}$ \\
   &  & &  &   & \\
  Red &  $0.0$ & $11.9\pm1.8\times10^{-3}$ & $0.13\pm0.15$ & $0.00\pm0.58$ & $-0.19_{-1.08}^{+1.09}$ \\
  Red &  $0.5$ & $11.3\pm2.1\times10^{-3}$ & $0.19\pm0.16$ & $0.05\pm0.60$ & $-0.14_{-1.10}^{+1.12}$ \\
  Red &  $1.0$ & $9.3\pm2.5\times10^{-3}$ & $0.28\pm0.18$ & $0.25\pm0.70$ & $0.20_{-1.31}^{+1.34}$ \\
  Red &  $1.5$ & $7.2\pm3.1\times10^{-3}$ & $0.40\pm0.22$ & $0.61\pm0.86$ & $0.87_{-1.63}^{+1.67}$ \\
  Red &  $2.0$ & $5.2\pm4.0\times10^{-3}$ & $0.54\pm0.27$ & $1.09\pm1.07$ & $1.82_{-2.08}^{+2.12}$ \\
   &   &   &  &  &\\
  Blue &  $0.0$ & $1.5\pm1.4\times10^{-3}$ & $-0.16_{-0.19}^{+0.18}$ & $-0.56\pm0.68$  & $-1.24_{-1.65}^{+1.62}$ \\
  Blue &  $0.5$ & $2.0\pm1.6\times10^{-3}$ & $-0.25\pm0.19$ & $-0.75\pm0.70$  & $-1.62_{-1.72}^{+1.69}$ \\
  Blue &  $1.0$ & $2.3\pm1.9\times10^{-3}$ & $-0.35_{-0.22}^{+0.21}$ & $-1.01\pm0.81$  & $-2.17_{-2.03}^{+1.97}$ \\
  Blue &  $1.5$ & $2.5\pm2.3\times10^{-3}$ & $-0.45\pm0.26$ & $-1.24\pm0.96$  & $-2.67_{-2.44}^{+2.36}$ \\
  Blue &  $2.0$ & $2.5\pm2.7\times10^{-3}$ & $-0.53_{-0.32}^{+0.31}$ & $-1.44\pm1.17$  & $-3.06_{-2.95}^{+2.85}$ \\
  \hline \\
  \end{tabular}
  \label{tab_fh}
\end{table*}  

\subsection{Environmental dependence}
\begin{figure}
  \resizebox{\hsize}{!}{\includegraphics{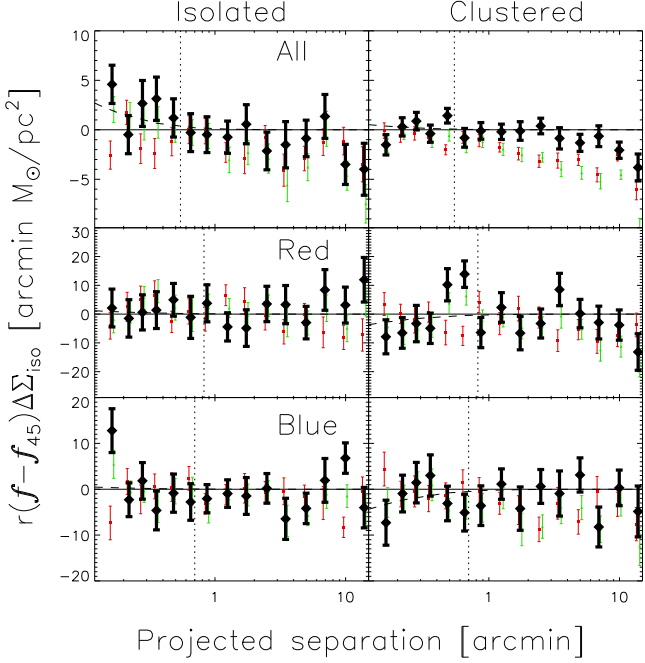}}
  \caption{Anisotropic shear signal multiplied with the projected separation in arcmin around the `all' sample (top panels), the `red' sample (middle panels) and the `blue' sample (bottom panels), for the isolated lenses on the left-hand side, and the clustered lenses on the right-hand side (using $w\propto e^{1.0}$). The green triangles denote $f \Delta \Sigma_{\rm{iso}}$, the red squares $f_{45} \Delta \Sigma_{\rm{iso}}$, and the black diamonds $(f-f_{45}) \Delta \Sigma_{\rm{iso}}$. The vertical dotted lines indicate the virial radius at the mean lens redshift, the dashed lines indicate the best fit elliptical NFW profiles. Please note the different scalings of the vertical axes. For the `all' sample, the shear anisotropy is larger for the isolated sample; for the `red' and `blue' sample, the shear anisotropy is more negative on small scales for the clustered samples, but the differences are not significant when averaged within the virial radius. }
  \label{plot_shearani_envir}
\end{figure}

\hspace{4mm} To study whether the lensing anisotropy depends on the lens environment we measure the signal for the isolated and clustered lens sample. In Figure \ref{plot_shearani_envir} we show $(f-f_{45})\Delta\Sigma_{\rm{iso}}$ for the $w\propto e^{1.0}$ bin, which is the same weight as used in \citet{Mandelbaum_ell06} and hence enables a direct comparison. 
We determine $\langle f-f_{45} \rangle$ and fit $f_{\rm{h}}$ for the elliptical density profiles, and show the results in Table \ref{tab_fh_envir}. \\
\indent The lensing anisotropy for the isolated `all' lenses is positive, and the values of $(f-f_{45}) \Delta \Sigma_{\rm{iso}}$ and $f_h$ are larger than those of the clustered sample by almost $\sim$2$\sigma$. For the `red' and the `blue' lenses, we find that on small scales, the lensing anisotropy is more negative for the clustered sample. When we average the signals within the virial radius, or fit the elliptical density profiles, we find that this difference is not statistically significant.   \\
\begin{table*}
  \caption{Similar to Table \ref{tab_fh} for the isolated and clustered lenses.}   
  \centering
  \begin{tabular}{c c c c c c c } 
  \hline
  Sample & $\alpha$ & Environment & $\langle f-f_{45} \rangle$ & $f_{\rm{h}}$(SIE)  & $f_{\rm{h}}$(NFW) \\
   &  &  &  &   & \\
  \hline\hline  \\
  All & 1.0 & isolated & $0.51_{-0.25}^{+0.26}$ & $2.35\pm1.00$  & $4.73_{-2.05}^{+2.17}$ \\
  All & 1.0 & clustered & $0.18\pm0.13$ & $0.40\pm0.51$ & $0.90_{-1.15}^{+1.17}$ \\
   &  &  &  &   & \\
  Red & 1.0 & isolated &  $0.11\pm0.25$ & $0.49\pm0.97$  & $0.40_{-1.90}^{+1.96}$  \\
  Red & 1.0 & clustered & $0.47\pm0.26$ & $0.46\pm1.05$  & $-1.23_{-1.83}^{+1.90}$ \\
   &  &  &  &   &\\
  Blue & 1.0 & isolated & $-0.18\pm0.30$ & $0.05\pm1.06$ & $0.38_{-2.45}^{+2.32}$ \\
  Blue & 1.0 & clustered & $-0.43_{-0.35}^{+0.34}$ & $-1.54\pm1.33$  & $-3.46_{-3.45}^{+3.26}$ \\  
  \hline \\
  \end{tabular}
  \label{tab_fh_envir}
\end{table*}  

\subsection{Interpretation \label{sec_int}}
\hspace{4mm} The shear anisotropy measurements provide weak support that the average galaxy is preferentially aligned with its triaxial dark matter host. The significance of the detection for the `all' sample does not depend on how we weigh the measurement with the observed galaxy ellipticity, which indicates that more elliptical galaxies do not reside in, or are better aligned with, more elliptical dark matter haloes. We find that the errors on $\langle f-f_{45} \rangle$ increase for larger values of the exponents of the weights, because the effective number of lenses decreases: round lenses barely contribute to the signal for the largest exponents. Furthermore, we find that the shear anisotropy signal of isolated galaxies is stronger than that of clustered galaxies: the difference is almost 2$\sigma$. One possible explanation is that the dark matter haloes of isolated galaxies are less subject to stripping, and may preserve their original shapes. Clustered galaxies, on the other hand, may lose more of their dark matter, particularly in the outer regions. An alternative explanation is that the fraction of lenses in the clustered sample that are satellites is larger. Since the host halo in which satellite galaxies are embedded dominates the lensing signal, and since the orientation of the major axis of the satellites and the major axis of the central galaxy are expected to not be strongly aligned, this would also lead to a reduction of the anisotropy of the lensing signal. \\
\indent The average shear anisotropy of both the `red' and `blue' lenses is consistent with zero. We find no significant differences between the clustered and isolated sample, because we lack precision. For the `red' lenses the increase of the detection significance of $\langle f-f_{45}\rangle$ to $\sim$2$\sigma$ by increasing the exponent of the lens ellipticity in the weights could be interpreted as an indication that red galaxies with larger ellipticities preferentially reside in more elliptical dark matter haloes, and/or are better aligned with them. However, Figure \ref{plot_shearani} reveals that the detection is caused by the two measurement points near the virial radius, and not by the measurements at the smallest scales where the signal should be highest if the galaxies are aligned with their triaxial dark matter haloes. This is reflected by the best-fit values of $f_h$ for the elliptical NFW profile, which are consistent with zero for all exponents. It seems therefore unlikely that the detection is caused by triaxial dark matter haloes. For the `blue' lenses, we find that for the largest exponents the anisotropy is marginally negative, which is suggestive of an anti-alignment between the galaxy and the dark matter halo.\\
\indent The negative shear anisotropy at small projected separations for the `red' lens sample occurs at the same scale where the contamination is highest. Since we observed in Section \ref{sec_contam} that the distribution of satellites is anisotropic, the observed signal could also be caused by a radial alignment of physically associated galaxies in the source sample. We estimate the value of the average tangential intrinsic alignment that would produce the signal we observe in Appendix \ref{ap_ia}. We find that we require a value that is roughly ten times larger then the results from \citet{Hirata04}. Conversely, the expected impact of intrinsic alignments is about a factor ten smaller than the signal we observe, hence it is unlikely that our measurements are significantly affected. Hence the negative shear anisotropy on small scales is unlikely caused by intrinsic alignments. \\
\indent Currently, weak lensing constraints on the ellipticity of dark matter haloes have been presented in \citet{Hoekstra04}, \citet{Mandelbaum_ell06} and \citet{Parker07}. We already compared our results with those from \citet{Parker07} in Section \ref{sec_rat}. \citet{Mandelbaum_ell06} measured the weak lensing anisotropy around $2\times10^6$ lenses with photometric redshifts in the SDSS. On average, they find $f_h=-0.06\pm0.19$ for the red galaxies, and $f_h=-1.1\pm0.6$ for the blue galaxies on scales 20-300 $h^{-1}$kpc for an SIE profile. Fitting an elliptical NFW profile yields $f_h=0.60\pm 0.38$ and $f_h=-1.4_{-2.0}^{+1.7}$ for the red and blue lenses, respectively. Separating the lenses in luminosity bins, they find that the lensing signal is consistent with zero for most of their bins. Only for the brightest red lenses a detection is reported, also at the $\sim$2 sigma level. A detailed comparison between the results is complicated due to differences in the lens selections. Our `red' lens samples can be best compared with the L5 and L6 red sample from \citet{Mandelbaum_ell06}, as these are most similar in absolute magnitudes; \citet{Mandelbaum_ell06} find $\langle f-f_{45} \rangle=0.08\pm0.08$ and $\langle f-f_{45} \rangle=0.29\pm0.12$ for L5 and L6, which agrees well with the $\langle f-f_{45} \rangle=0.28\pm0.18$ that we found. For the elliptical NFW profile, \citet{Mandelbaum_ell06} found $f_h=0.4\pm0.57$ and $f_h=1.7\pm0.7$ for the red L5 and L6 sample, and the average of those values is roughly within 1$\sigma$ of our best fit value $0.20^{+1.34}_{-1.31}$. Our `blue' lens sample covers a broad range in luminosity, and roughly corresponds to the blue lens bins L3 to L5 of \citet{Mandelbaum_ell06}. For these bins, \citet{Mandelbaum_ell06} finds $\langle f-f_{45}\rangle=-0.29^{+0.26}_{-0.27}$, $-0.36^{+0.25}_{-0.26}$ and $-0.27\pm0.28$ respectively, which agrees well with $-0.35^{+0.21}_{-0.22}$. In conclusion, we find that our results and the results from \citet{Mandelbaum_ell06} are consistent.  \\
\indent In an earlier work, \citet{Hoekstra04} used 45.5 deg$^2$ of the RCS \citep{Gladders05} to measure the lensing anisotropy around $1.2\times10^5$ lenses selected with $19.5<R_C<21$. A $\sim$2$\sigma$ detection of $f_h=0.77^{+0.18}_{-0.21}$ was obtained by fitting a TIS using a maximum likelihood method. This result appears to be significantly different from ours, and from \citet{Mandelbaum_ell06}. In the latter, various reasons are given why the results could differ, which also apply to us: most importantly, the lens samples are very different, and the method of analysis differs. Possibly, the maximum likelihood method is a better estimator of the shear anisotropy, as it takes into account the positions and relative orientations of the lens galaxies. The result from a maximum likelihood fit, however, is difficult to interpret as it is not clear how multiple deflections affect the measurement, and because the radial dependence of the signal cannot be visualized, which makes it difficult to identify residual systematics. \\

\begin{figure}
  \resizebox{\hsize}{!}{\includegraphics[width=14cm]{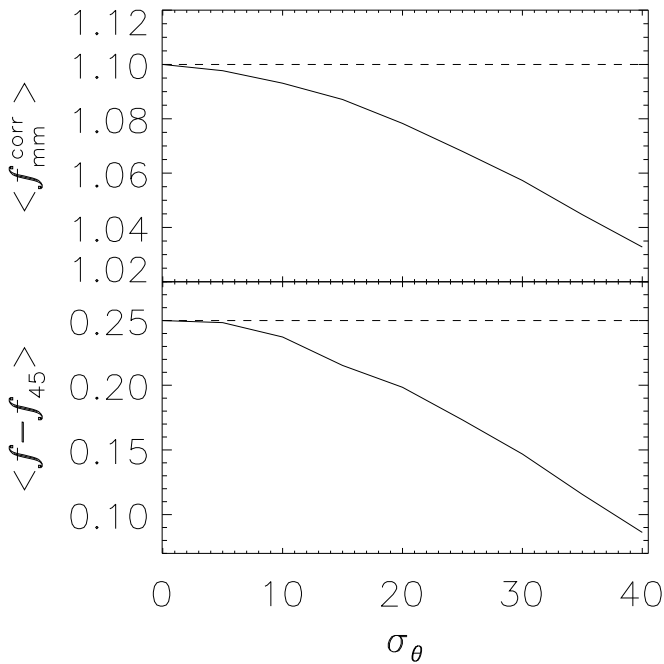}}
  \caption{Anisotropy of the lensing signal averaged between 1 and 200 arcsec, for a Gaussian distributed position angle difference between the dark matter and the light distribution with zero mean and width $\sigma_{\theta}$. Large values for $\sigma_{\theta}$ as have been reported in the literature lead to significant reductions in the shear anisotropy. }
  \label{plot_modDTheta}
\end{figure}

\indent Converting the average shear anisotropy into the average projected ellipticity of dark matter haloes is complicated. The simulations in this work have shown that multiple deflections and clustering can have some impact on the lensing anisotropy. If we select the lens sample carefully as we did (in particular the `blue' and `red' lens sample that consist of isolated, elliptical, low-redshift, massive lenses), the impact of these complications is small. However, misalignments between the position angles of the galaxies and their dark matter haloes cause a reduction of the lensing anisotropy. Therefore, the observational constraints on the shear anisotropy from previous weak lensing analyses, as well as from this work, only provide lower limits on the average halo ellipticity. To constrain the halo ellipticity from the lensing anisotropy, we need to know how galaxies are orientated in their triaxial dark matter haloes. \\
\indent The orientation of galaxies in their dark matter haloes has received considerable attention in recent years. Studies of the distribution of satellite galaxies around centrals in the SDSS \citep{Wang08,Agustsson10} and in numerical simulations \citep{Kang07,Deason11}, studies based on the ellipticity correlation functions of galaxies \citep{Faltenbacher09,Okumura09}, and studies based on angular momentum considerations in numerical simulations \citep{Bett10,Hahn10} all point in a similar direction: on average, red galaxies are aligned with their dark matter hosts, but with a considerable scatter between the position angles with a value in the range $\sim$20-40 degrees. The dispersion for blue galaxies is even larger. \\
\indent Scatter between the position angles irrevocably leads to a reduction in the anisotropy of the lensing signal; with lensing, we only measure the component of the dark matter halo ellipticity that is aligned with the lens light. To study the magnitude of the reduction, we use the simulations from the previous section, where the position angle of the dark matter halo is given by the position angle of the light, plus a Gaussian with zero mean and a certain width. In Figure \ref{plot_modDTheta}, we show the average $f_{\rm{mm}}^{\rm{corr}}$ and $\langle f-f_{45} \rangle$ as a function of $\sigma_{\theta}$, the dispersion of the position angle distribution. We find that a $\sigma_{\theta}$ in the range between 20 to 40 degrees leads to a reduction in the anisotropy of $\sim$25-65\%. This result indicates that the scatter in the relative position angle significantly dilutes the shear anisotropy measurements. \\
\indent Qualitatively similar results have been obtained by \citet{Bett11}, who used the Millennium simulation \citep{Springel05} in combination with semi-analytic galaxy formation models to predict the stacked projected axis ratio, $q$, of large numbers of haloes. Various alignment models were tested, and different methods used to measure the halo shapes. For most scenarios, $q$ turned very close to unity, implying an nearly isotropic shear signal. \\
\indent In reality, the alignment between the position angle of the galaxy and the dark matter may be scale dependent. Besides that, the dark matter haloes are not rigid once formed; their shapes continue to evolve in a way that depends on, amongst others, their formation history and environment; the galaxies in the centre of the haloes may evolve differently. Furthermore, the central galaxy is expected to sphericalise the dark matter haloes \citep[e.g.][]{Kazantzidis10,Abadi10,Machado10}. Indications already exist that the alignment depends on lens luminosity \citep[][]{Faltenbacher09,Mandelbaum_ell06} and environment \citep{Hahn10}; it could depend on other lens characteristics (e.g. lens ellipticity, redshift) as well. Finally, once more, multiple deflections and clustering are expected to have some impact on the anisotropy of the lensing signal. All these effects have to be accurately modeled and well understood before we can interpret any measured anisotropy in terms of the average property of dark matter haloes. Note that the study of intrinsic alignments of galaxies are similarly affected, and require this knowledge as well for a correct interpretation. \\

\indent Our results underline the need for photometric redshifts - and consequently luminosities - for the lenses. Without photometric redshifts, we can only select lenses based on their colours and magnitudes. To achieve sufficient signal-to-noise in order to obtain competitive constraints on the average halo ellipticity, we have to select large numbers of galaxies that cover a broad range of luminosities and redshifts. If the average halo ellipticity depends on the luminosity of a galaxy (as the results from Mandelbaum et al. 2006a suggest), the signal-to-noise of the shear anisotropy measurements decrease, and may even average out in the worst case scenario. If luminosities are available, we can not only select lenses in narrow luminosity ranges, but also weigh the lensing measurement with luminosity, which improves the signal-to-noise of the lensing measurement. The lack of photometric redshifts also forces us to stack the lensing signal as a function of angular separation, rather than physical, which decreases the signal-to-noise as well. This is particular disadvantageous as the shear anisotropy signal of an elliptical NFW profile drops very rapidly with increasing radius, and the signal may be smeared out and become undetectable. We note that a preliminary photometric redshift catalogue exists for the RCS2 for the area that has also been observed in the $i'$-band, but it only covers the redshift range $z>0.4$ due to the absence of observations in the $u$-band, which limits its usefulness for this study. \\

\indent A new technique has recently been proposed to improve halo ellipticity measurements: the use of a higher order distortion of lensing known as flexion \citep{Hawken09,Er11a,Er11b,Er11c}. Although the measurement of the flexion signal is difficult for galaxy-scale potentials, the first positive detections have already been reported \citep{Velander11}. Using mock simulations of clusters with SIE and elliptical NFW profiles, \citet{Er11c} find that flexion is more sensitive to the halo ellipticity than the shear; this may be true as well for stacked galaxy potentials. Furthermore, the systematic errors in flexion measurements differ from those in shear. Hence we anticipate that additional useful constraints can be obtained with flexion. 

\section{Conclusion \label{sec_concl5}}
\hspace{4mm} We present measurements of the anisotropy of the weak lensing signal around galaxies using data from the Red-sequence Cluster Survey 2 (RCS2). We define three lens samples: the `all' sample contains all galaxies in the range $19<m_{r'}<21.5$, whereas the `red' and `blue' samples are dominated by massive low-redshift early-type and late-type galaxies, respectively. To study the environmental dependence of the lensing signal, we also subdivide each lens sample into an isolated and clustered part, and analyse them separately. \\
\indent We address the impact of several complications on the shear anisotropy measurements, including residual PSF systematics in the shape catalogues, multiple deflections, the clustering of lenses, and correlations between their intrinsic shapes. We run a set of idealised simulations to estimate the impact these might have on real data, and find them to be small, but not entirely negligible. We demonstrate that the impact of these complications can be reduced by a careful selection of the lens sample, i.e. low-redshift, massive and elliptical galaxies, as has been done in this work.  \\
\indent We also measure the distribution of physically associated galaxies around the lens samples. We find that these satellites predominantly reside near the major axis of the lenses. The results of the `red' sample are in good agreement with previously reported values, whilst the constraints of the `all' and `blue' sample cannot be easily compared as they consist of a mixture of early-type and late-type galaxies. \\
\indent The shear anisotropy is quantified by the anisotropy of the galaxy-mass cross-correlation function, $\langle f-f_{45}\rangle$, and by the ratio of the projected dark matter halo ellipticity and the observed galaxy ellipticity, $f_h$.  For the `all' sample we find that $\langle f-f_{45}\rangle=0.23\pm0.12$, and $f_h=1.50_{-1.01}^{+1.03}$ for an elliptical NFW profile, which for a mean lens ellipticity of 0.25 corresponds to a projected halo ellipticity of $e_h=0.38_{-0.25}^{+0.26}$ if the halo and the lens are perfectly aligned. Note that various studies indicate that this may not be the case. These constraints provide weak support that galaxies are embedded in, and preferentially aligned with, triaxial dark matter haloes. For isolated galaxies, the average shear anisotropy is larger than for clustered galaxies; for elliptical NFW profiles, we find $f_h=4.73_{-2.05}^{+2.17}$ and $f_h=0.90_{-1.15}^{+1.17}$, respectively. The decrease of the lensing anisotropy signal around clustered galaxies may be due to the stripping of dark matter haloes in dense environments. \\
\indent We do not detect a significant shear anisotropy for the average `red' lens. The shear anisotropy for the most elliptical `red' galaxies is marginally positive, but the pattern of the signal suggests that this is not the result of an alignment between the dark matter haloes and the galaxies. For the `blue' lenses, we find that the shear is marginally negative, with slightly increased significance for the most elliptical galaxies, suggesting an anti-alignment between the galaxy and the dark matter. Our measurements highlight the need for (photometric) redshifts in lensing studies. In order to reach sufficient signal-to-noise that enable competitive constraints on the shear anisotropy, we have to stack large numbers of galaxies that span a broad range in luminosities and redshifts. This smears out the shear anisotropy, and in the worst case the anisotropy might even average out. \\

\paragraph{Acknowledgements \\ \\} 
We would like to thank Elisabetta Semboloni for useful discussions and suggestions on the error estimation of the shear anisotropy measurements, Peter Schneider for valuable comments on the manuscript, and Rachel Mandelbaum for providing the table of the shear anisotropy of an elliptical NFW profile as well as for valuable comments on this work. HH and EvU acknowledge support from a Marie Curie International Reintegration Grant. HH is also supported by a VIDI grant from the Nederlandse Organisatie voor Wetenschappelijk Onderzoek (NWO). TS acknowledges support from NSF through grant AST-0444059-001, and the Smithsonian Astrophysics Observatory through grant GO0-11147A. MDG thanks the Research Corporation for support via a Cottrell Scholars Award. The RCS2 project is supported in part by grants to HKCY from the Canada Research Chairs program and the Natural Science and Engineering Research Council of Canada. \\
\indent This work is based on observations obtained with MegaPrime/MegaCam, a joint project of CFHT and CEA/DAPNIA, at the Canada-France-Hawaii Telescope (CFHT) which is operated by the National Research Council (NRC) of Canada, the Institute National des Sciences de l'Univers of the Centre National de la Recherche Scientifique of France, and the University of Hawaii. We used the facilities of the Canadian Astronomy Data Centre operated by the NRC with the support of the Canadian Space Agency. 

\bibliographystyle{aa}


\begin{appendix}
\section{Lens selection \label{ap_lenssel}}
\begin{figure*}
  \resizebox{\hsize}{!}{\includegraphics[width=14cm,angle=-90]{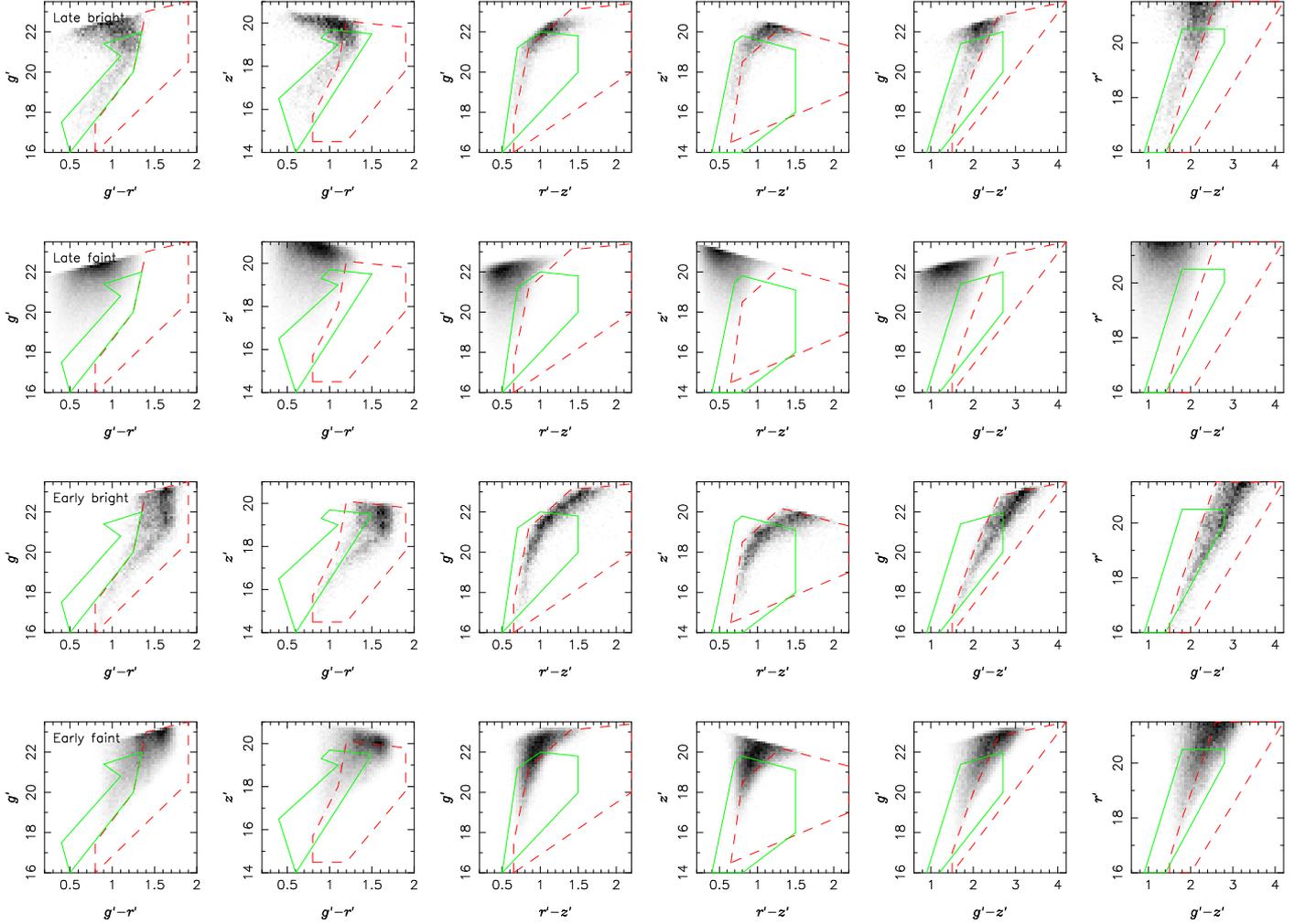}}
  \caption{Colour-magnitude diagrams that are used to define the lens sample selection. In gray we show the locations of the galaxies with $m_{r'}<21.5$ from the W1  photometric redshift catalogue from the CFHTLenS collaboration \citep{Hildebrandt11}. In the top row we show the late-type galaxies with an absolute magnitude $M_{r'}<-22.5$, in the second row we show the late-types with  $M_{r'}>-22.5$, and in the third and fourth row we show the early-types with  $M_{r'}<-22.5$ and $M_{r'}>-22.5$, respectively. The coloured boxes illustrate the lens sample selection criteria; all galaxies that reside in all the red dashed (green solid) boxes form the `red' (`blue') lens sample. We find that the `red' lens sample contains very few late-types, as they are excluded in the $g'-r'$ versus $g'$ diagram (first column). Also, most faint early-type galaxies are excluded from this sample as well (see, e.g., the third and fourth columns). For the `blue' lens sample, we cannot define selection criteria that exclusively select bright late-type galaxies, and this sample therefore also contains a number of faint late-type and early-type galaxies.  }
  \label{plot_colmag}
\end{figure*}
We define our `red' and `blue' lens samples using the W1  photometric redshift catalogue from the CFHTLenS collaboration \citep{Hildebrandt11}. This catalogue contains the apparent magnitudes of galaxies observed in the same filters as in the RCS2, but also their photometric redshifts and absolute magnitudes. We split the sample in quiescent (typically early-type) and star-forming (typically late-type) galaxies according to their photometric type identifier $T_{\rm{BPZ}}$. $T_{\rm{BPZ}}$ corresponds to the galaxy type of the best-fit template (E/S0, Sbc, Scd, Im, SB3, SB2, with $T_{\rm{BPZ}}$ values from 1.0 to 6.0). We select the early-types and late-types with $T_{\rm{BPZ}}<1.5$ and $T_{\rm{BPZ}}>1.5$ respectively. These selections are further divided into a bright and faint sample using  $M_{r'}<-22.5$ and $M_{r'}>-22.5$. For these four samples, we plot the colours as a function of magnitude in Figure \ref{plot_colmag}. We define selection boxes of the `red' and `blue' lens sample, which are aimed at selecting the bright early-types and late-types. Figure \ref{plot_colmag} shows that the `red' lens sample contains almost no late-type galaxies, nor faint early-types. Unfortunately, for the `blue' sample we cannot exclusively select luminous late-type galaxies, and the sample also contains faint late-type and early-type galaxies. These selection criteria are applied to the catalogues of the RCS2 to select the `red' and `blue' lens sample.  \\
\indent To study how well we can separate early-types from late-types, we compare our selection to previously employed separation criteria. We find that 98\% of the `red' lenses, 57\% of the `blue' lenses and 28\% of the `all' lenses have a photometric type $T_{\rm{BPZ}}<1.5$, and therefore have a spectral energy distributions similar to red elliptical galaxies. The `red' sample therefore barely contains late-type galaxies. The majority of the `blue' sample are actually faint red early-type galaxies, as can be seen from Figure \ref{plot_colmag}. The lensing signal is dominated, however, by the massive late-type galaxies. The faint lenses mainly add noise. The purity of the `blue' sample could be improved by shifting the selection boxes to bluer colours, but this at the expense of removing the majority of massive late-type lenses. Finally, the majority of the `all' sample are blue and not very massive late-type galaxies. As a check, we also compare to the $u-r$ colour selection criterion, which has been used in \citet{Mandelbaum_ell06}. In this work, galaxies with an SDSS $u-r>2.22$ model colour are selected for the red sample, whilst galaxies with $u-r<2.22$ are selected as blue galaxies. When we select galaxies based on identical $u'-r'$ criteria (hence ignoring small differences between the filters), we find very similar results: 99\% of the `red' lenses, 58\% of the `blue' lenses and 30\% of the `all' lenses have a colour $u'-r'>2.22$ and are red. 


\section{Environment selection \label{ap_envir}}
\begin{figure*}
  \resizebox{\hsize}{!}{\includegraphics{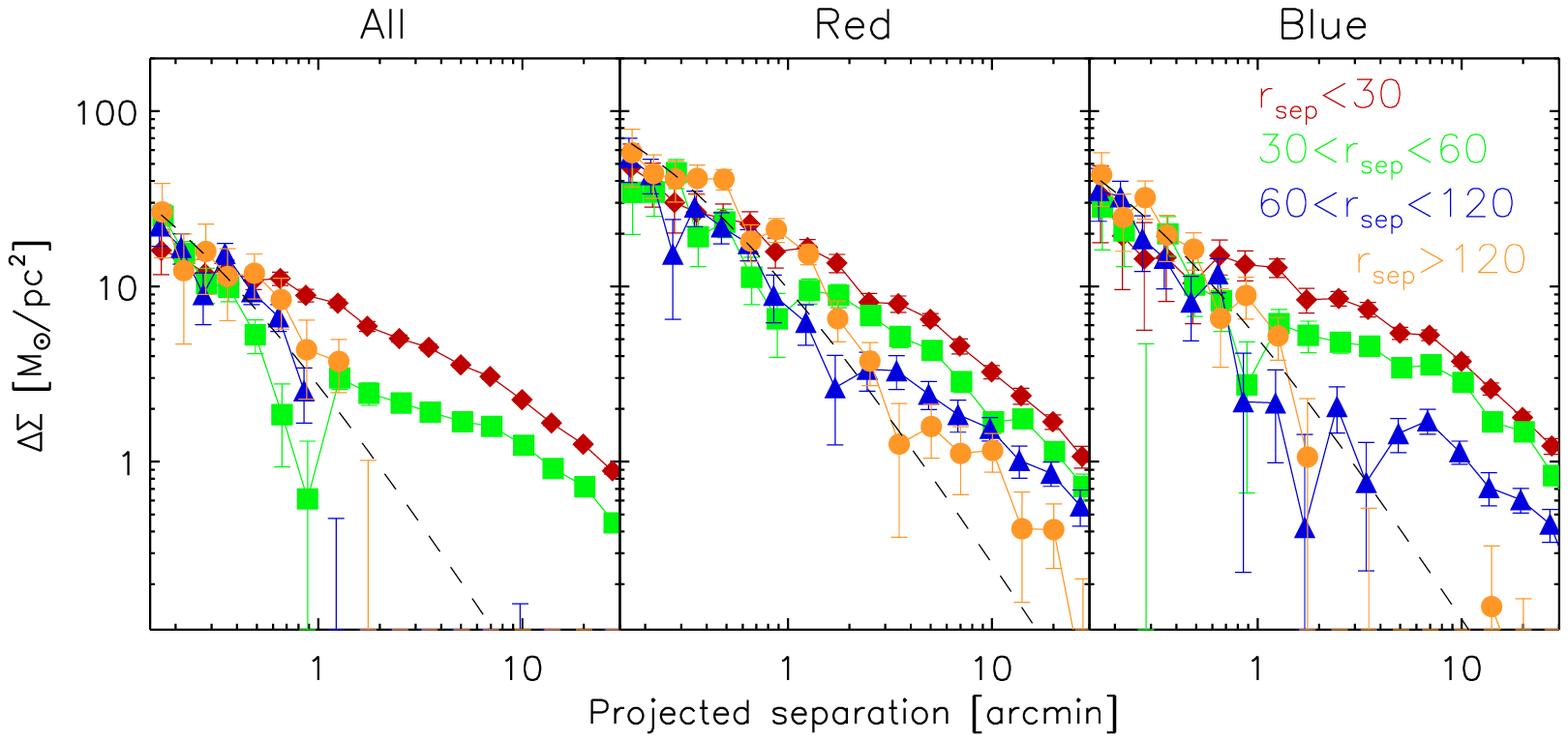}}
  \caption{The lensing signal as a function of projected separation for the three lens samples that have been divided according to the minimum distance to a brighter galaxy, as indicated in the top right panel. The red diamonds indicate the lensing signal of those lenses that have a brighter neighbour within 30 arcsecs, the green squares the signal of the lenses that have a brighter neighbour between 30 and 60 arcsecs, the blue triangles the signal of the lenses that have a brighter neighbour between 60 and 120 arcsecs, and the orange circles the signal of the lenses that have no brighter neighbour within 2 arcmins. The dashed lines show the best fit NFW profiles fitted to the lensing signal of the total sample between 50 and 500 kpc using the mean lens redshift. The lensing signal on scales $<$0.5 arcmin is roughly similar for all lens samples, but the lensing signal on larger scales clearly decreases for increasingly isolated galaxies.}
  \label{plot_envirsel}
\end{figure*}

We subdivide the lenses in a clustered and an isolated sample, depending on whether or not the lens has a neighbour within a certain projected radius range that has a lower apparent magnitude than the lens. To determine which radius effectively separates the lenses into a low-density and a high-density sample, we compare the lensing signals for four lens selections: those that have a brighter neighbour within 30 arcsecs, those with a brighter neighbour between 30 arcsecs and 1 arcmin, those with a brighter neighbour between 1 arcmin and 2 arcmins, and those that are the brightest object within 2 arcmins. We show the lensing signal of the four selections in Figure \ref{plot_envirsel}. 
We find that the lensing signal at scales smaller than $\sim$0.5 arcmin does not change much using the different isolation criteria, which demonstrates that we select haloes of similar mass. However, the large-scale signal decreases significantly for an increasing minimum separation to a brighter neighbour. The galaxies with no brighter object within 1 arcmin are selected for the isolated sample, the other galaxies are selected for the clustered sample. We could in principle select a more clearly distinguished sample of isolated and clustered lenses, e.g. by only selecting those galaxies with no brighter neighbour within 2 arcmins, and those with a brighter neighbour within 30 arcsecs. However, this would reduce the signal-to-noise of the lensing measurements such that no useful constraints could be obtained on the average halo ellipticity, and we therefore choose not to. \\
\indent Note that the dependence of the large-scale lensing signal on the distance to a brighter neighbour is partly caused by differences in the lens environment, and partly by differences of the projected densities along the line-of-sight (LOS). For the clustered lenses, the LOS projections do not simply add random noise, but systematically increase the large-scale lensing signal because of the neighbour selection. Similarly, for the isolated galaxies, the underdense LOS leads to a decrease of the lensing signal on large scales. For the purpose of this work this is not important, since our main focus is to measure the anisotropy of the lensing signal. However, it should be kept in mind in the interpretation of the results. \\

\section{Average ratios and their errors \label{ap_asy}}
To calculate $f_{\rm{mm}}$, $f^{\rm{corr}}_{\rm{mm}}$ and $(f-f_{45})$ we have to determine the ratio of two variables. Let us call these variables $x$ and $y$, and $m=x/y$ the ratio we are interested in. If $x$ and $y$ are independent and have a Gaussian distribution with means $\mu_x$ and $\mu_y$ and widths $\sigma_x$ and $\sigma_y$ we can compute the probability distribution of $m$ using \citep{Hinkley69}
\begin{equation}\begin{split}
p(m)=\frac{b(m)c(m)}{\sqrt{2\pi}a^3(m)\sigma_x\sigma_y}\Big[2\Phi\bigg(\frac{b(m)}{a(m)} \bigg) \Big] + \\
\frac{1}{\pi a^2(m)\sigma_x\sigma_y}e^{-\frac{1}{2}\big(\frac{\mu^2_x}{\sigma^2_x}+\frac{\mu^2_y}{\sigma^2_y} \big)},
\label{eq_hink}
\end{split}\end{equation}
where
\begin{equation}
a(m)=\sqrt{\frac{1}{\sigma^2_x}m^2+\frac{1}{\sigma^2_y}},
\end{equation}
\begin{equation}
b(m)=\frac{\mu_x}{\sigma^2_x}m+\frac{\mu_y}{\sigma^2_y},
\end{equation}
\begin{equation}
c(m)=e^{\frac{1}{2}\frac{b^2(m)}{a^2(m)}-\frac{1}{2}\big(\frac{\mu^2_x}{\sigma^2_x}+\frac{\mu^2_y}{\sigma^2_y} \big)},
\end{equation}
and
\begin{equation}
\Phi(m)=\int_{-\infty}^{m}\frac{1}{\sqrt{2\pi}}e^{-\frac{1}{2}u^2}du.
\end{equation}
By integrating $p(m)$ we can determine the median, and the 68\% confidence intervals. \\
\indent Next, we want to combine various measures of $m_i$ into one average. We follow the approach described in \citet{Mandelbaum_ell06}. It relies on the use of a slightly different variable, i.e. $y_i-mx_i$, which is again a random Gaussian variable. If the shear ratio is constant over the range of interest, its value can be determined with
\begin{equation}
\frac{-Z}{\sqrt{\sum w_i}} < \frac{\sum w_i(y_i-mx_i)}{\sum w_i} < \frac{Z}{\sqrt{\sum w_i}},
\label{eq_z}
\end{equation}
where $w_i=1/(\sigma^2_{y_i}+m^2\sigma^2_{x_i})$ and $\sigma^2_{y_i}$ and $\sigma^2_{x_i}$ the error on $y_i$ and $x_i$, respectively. $Z=0$ then gives the average ratio, whilst $Z=1$ gives the 68\% confidence intervals. Note that Equation (\ref{eq_z}) can also be used to determine the ratio of a single measurement, but we prefer the use of Equation (\ref{eq_hink}) as it directly provides the full probability distribution. \\

\section{Lens light contamination \label{ap_lenslight}}
\begin{figure}
  \resizebox{\hsize}{!}{\includegraphics{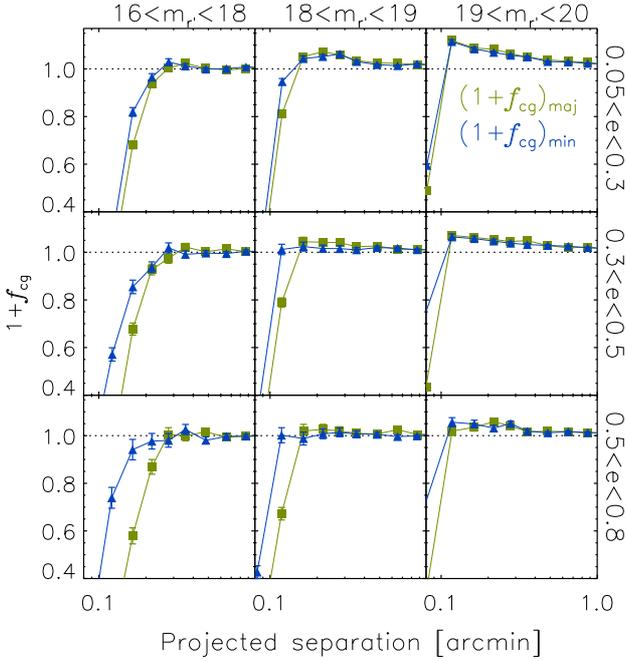}}
  \caption{The source number density along the major and minor axis close to bright foreground galaxies, indicated by green squares and blue triangles, respectively, selected on their apparent magnitudes (columns) and ellipticity (rows). On small scales, the deficiency of background galaxies along the major axis is larger for brighter and more elliptical galaxies. }
  \label{plot_odcontam}
\end{figure}

The light from bright and elliptical lenses changes the source number density along the major and minor axis differently on small projected separations close to the lens. This could bias the correction we make to account for physically associated galaxies in the source sample. We investigate the size of the effect by selecting all galaxies with $16<m_{r'}<18$, $18<m_{r'}<19$ and $19<m_{r'}<20$, and divide each selection in three ellipticity bins. For each of these samples, we measure the source number density in the major and minor axis quadrants, and show the results in Figure \ref{plot_odcontam}. On small scales, we find that the deficiency in the major axis quadrants is indeed significantly larger, and the difference increases for brighter and more elliptical galaxies. For projected separations larger than 0.2 arcmin, the difference is negligible, except for the brightest and most elliptical bins. Therefore, for galaxies with $m_{r'}<19$ we only use scales larger than 0.2 arcmin, whilst for galaxies with $m_{r'}>19$ we use scales larger than 0.1. \\

\section{Magnification \label{ap_magni}}
Magnification may also affect the shear anisotropy measurements. In this appendix, we address two potential contaminants: anisotropic magnification by the lens, and additional magnification by other foreground galaxies.
\subsection{Anisotropic magnification\label{ap_magni_ani}}
If dark matter haloes are triaxial, the magnification of background sources will have an azimuthal dependence. Hence if the galaxy and dark matter halo are aligned, part of the observed anisotropy in the distribution of source galaxies may be due to anisotropic magnification. The shear anisotropy measurements are corrected, however, assuming that the anisotropic distribution is solely due to physically associated galaxies. If anisotropic magnification has a strong effect on the source density, the correction might therefore be biased. To obtain an estimate of the impact of anisotropic magnification, we assume that the galaxy and the dark matter halo are perfectly aligned. Furthermore, we assume that at small scales the stacked density profile of the lenses is approximately described by an SIE, with a surface density given by \citep{Mandelbaum_ell06}
\begin{equation}
\kappa=\frac{4\pi\sigma^2}{c^2}\frac{D_lD_{ls}}{D_s}\frac{1}{2r} \times \bigg[1+\frac{e_h}{2}\cos(2\theta)\bigg].
\label{eq_kappasis}
\end{equation}
$\sigma$ is the average velocity dispersion, which we determine by fitting an SIS to the azimuthally averaged tangential shear within the virial radius, $D_{ls}/D_{s}$ is determined by integrating over the source redshift distribution, and the halo ellipticity $e_h$ is assumed to be equal to the mean galaxy ellipticity, as tabulated in Table \ref{tab_lenssamp} (i.e. assuming $f_{\rm{h}}=1$). We determine the magnification using $\mu=1+2\kappa$ at a projected separation of 10 arcsecs. The change in number density due to magnification in the major axis quadrant, $f^{\rm{mag}}_{\rm{B}}(r)$, is calculated with \citep{Narayan89}
\begin{equation}\begin{split}
f^{\rm{mag}}_{\rm{B}}(r)=\frac{2}{\pi}\int_{-\pi/4}^{\pi/4} d\theta N_{\rm{B}}(m,r,\theta)/N_0(m)dm \\
  =\frac{2}{\pi}\int_{-\pi/4}^{\pi/4} d\theta\mu_{\rm{B}}^{2.5s(m)-1}dm,
\end{split}\end{equation}
where $N_0(m)$ is the background galaxy number density, $N_{\rm{B}}(m,r,\theta)$ is the number density after magnification in the major axis quadrant, and $s(m)$ is the slope of the logarithmic galaxy number counts at magnitude $m$. The change in number density along the minor axis, $f^{\rm{mag}}_{\rm{A}}(r)$, is calculated similarly. We determine $s(m)$ using the photometric redshift catalogue of \citet{Ilbert06}, and find that it decreases from 0.34 at $m_{r'}=22$ to 0.22 at $m_{r'}=24$; we use the average value $s(m)=0.28$. Hence the number density of our source sample is diluted due to magnification, more strongly along the major axis of the lenses than along the minor axis as the magnification along the major axis is larger. For the `all' lens sample, we find that at 10 arcsecs, $f^{\rm{mag}}_{\rm{B}}=0.9969$ and $f^{\rm{mag}}_{\rm{A}}=0.9974$. For the `red' lens sample we find $f^{\rm{mag}}_{\rm{B}}=0.9927$ and $f^{\rm{mag}}_{\rm{A}}=0.9935$, and for the `blue' lens sample we find $f^{\rm{mag}}_{\rm{B}}=0.9946$ and $f^{\rm{mag}}_{\rm{A}}=0.9954$. Hence the effect of anisotropic magnification is expected to be very small, and does not cause the observed anisotropy of the excess source galaxy density ratio. \\

\subsection{Magnification by L2}
\begin{figure}
  \resizebox{\hsize}{!}{\includegraphics[width=14cm]{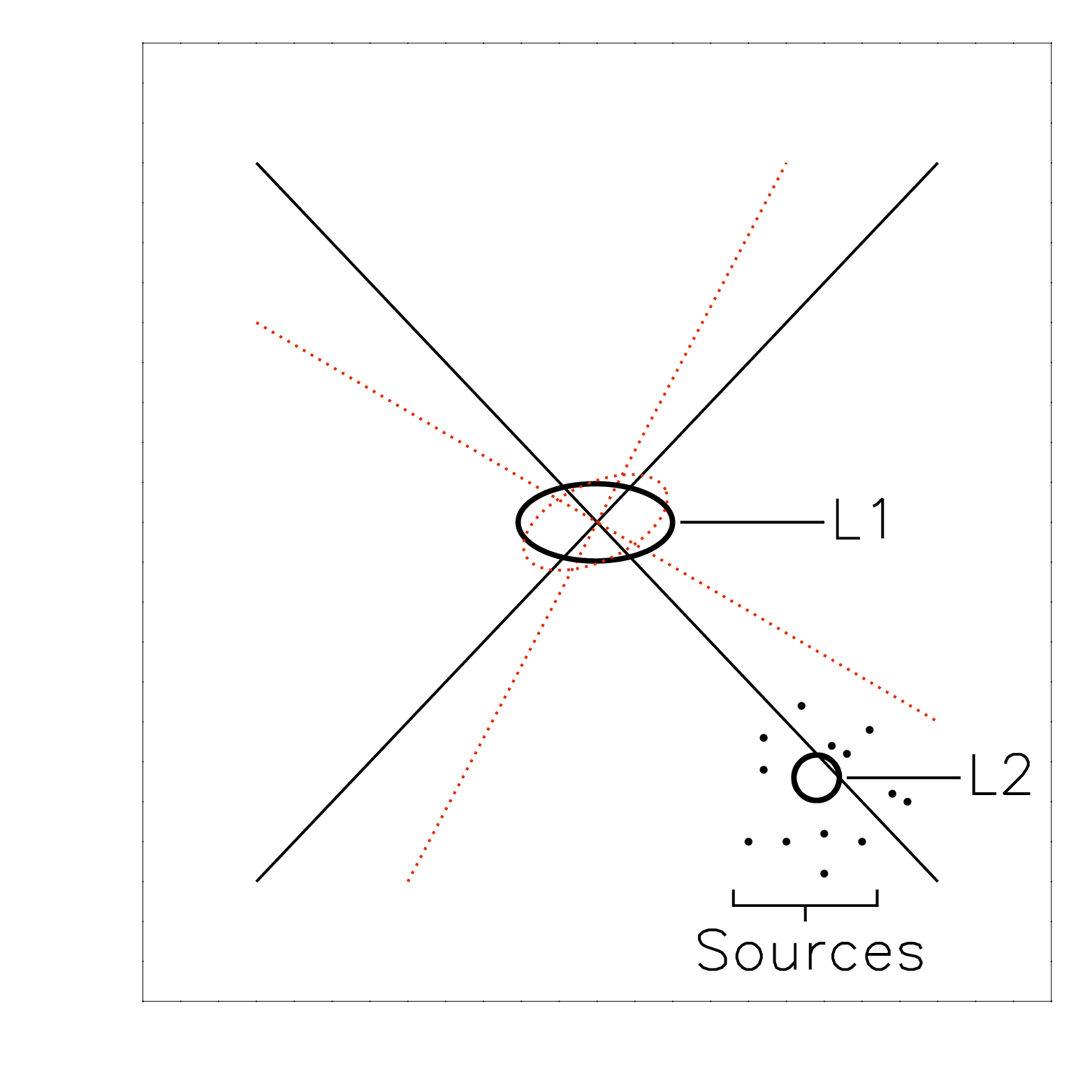}}
  \caption{Cartoon to illustrate a potential bias in the halo ellipticity measurement due to magnification. In the presence of a foreground galaxy L2, the quadrants where the tangential shear is averaged are rotated. L2 also magnifies the background sky, leading to an relative increase or decrease in number of sources in the minor axis quadrant. This potentially biases the measured shear ratio on small scales. }
  \label{plot_cartMagn}
\end{figure}

Foreground galaxies that lens both the lens galaxy and the sources also magnify the background sky. This causes a change in the source number density around the foreground galaxy. Consequently, the positions of the foreground galaxy and the background sources become related. This could potentially lead to a false shear anisotropy signal. We explain the effect using a cartoon in Figure \ref{plot_cartMagn}. \\
\indent In this figure, we measure the shear anisotropy around lens L1. The intrinsic shape of L1 is shown by the black ellipse, and the regions where we average the tangential shear in the absence of a foreground lens are indicated by the black lines. In the presence of a foreground lens L2 that is located close to one of the original quadrant's axes, the position angle of L1 changes, and the quadrants where we average the shear rotate, as indicated by the red dotted lines. L2 also locally magnifies the background sky, leading to either an increase or decrease in the source number density around L2. When we average the tangential shear in the rotated quadrants as would be done in observations, these magnified sources predominantly move into the minor axis quadrant (region {\tt A} in Figure \ref{plot_lensschematic}). If the source number density increases around L2, we find that we have more sources in the minor axis quadrants than in the major axis quadrants. Furthermore, the average shear changes, biasing the shear ratio high. If the source number density decreases instead, the opposite effect happens. \\
\indent The presence of this bias could be identified by comparing the number counts of the sources in the two quadrants for the radial bins close to the lens, as the effect is strongest on small scales. The effect is mixed, however, with the anisotropic distribution of satellite galaxies. However, we already found in Appendix \ref{ap_magni_ani} that magnification for our source sample is negligible. Therefore, we expect that this source of bias is small, and can be ignored with the current data. The effect may be measurable by selecting a sample of source galaxies that are clearly in the background, and whose number density slope is steep (e.g. Lyman-break galaxies), as these are the conditions favourable to magnification. \\

\section{Multiple deflections \label{ap_md}}

\begin{figure*}
  \resizebox{\hsize}{!}{\includegraphics[angle=270]{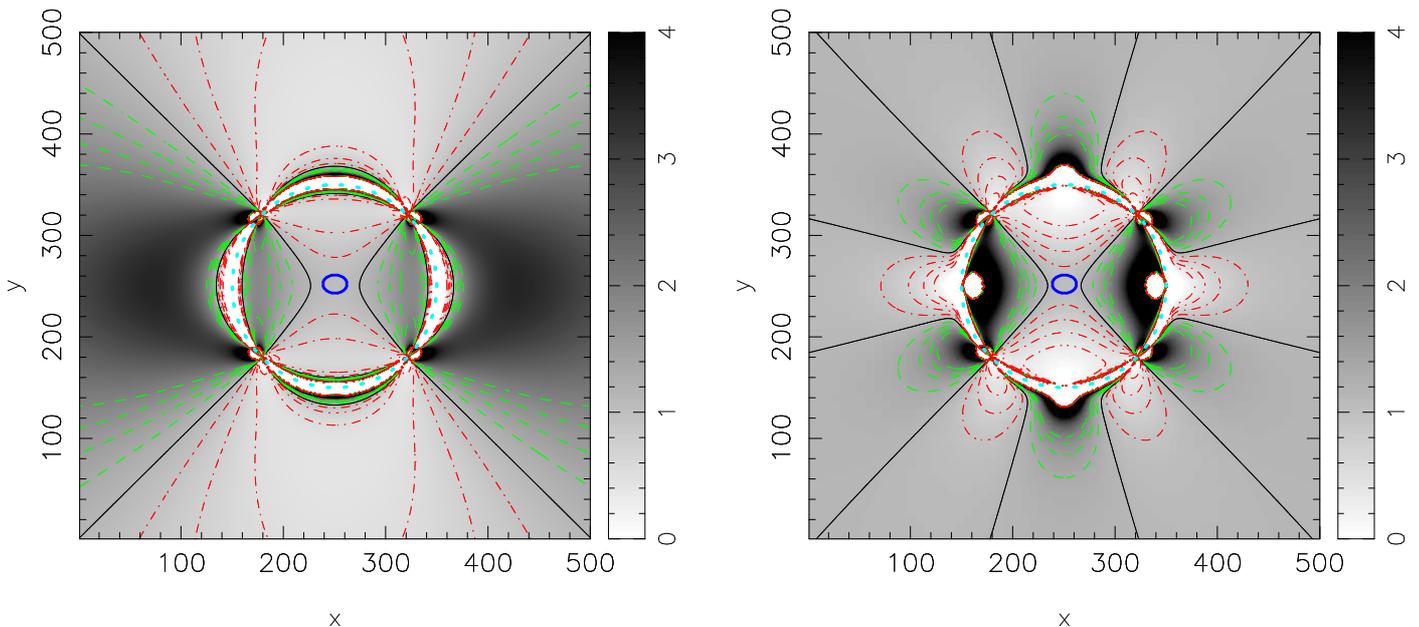}}
  \caption{Observed shear ratio of an elliptical lens L1 at $z=0.4$ for sources at a projected separation of 100 arcsecs, in the presence of an additional foreground galaxy L2 at $z=0.1$ that is located at ($x,y$). In the left-hand panel we show $\tilde{f}_{\rm{mm}}$, the ratio of the average shear along the major and minor axis. The location of L1 is indicated by the blue ellipse in the centre of the image, the location of the sources by the cyan dots. We find that depending on the position of L2, the shear ratio either increases or decreases. The black solid lines indicate the position of L2 where $\tilde{f}_{\rm{mm}}$ remains constant, the red dot-dashed (green dashed) lines indicate a decrease (increase) of $\tilde{f}_{\rm{mm}}$ of 0.2, 0.4 and 0.6, respectively. In the right-hand panel, we show $\tilde{f}_{\rm{mm}}^{\rm{corr}}$, the shear ratio corrected using the cross terms. For large L2-source separations, the induced shear on the lens and the sources is almost constant, and is effectively removed using the cross terms. For small separations, the residual is large as the applied shear from L2 varies strongly along the ring of sources. Note that both L1 and L2 are modeled by an SIS. }
  \label{plot_multdefl}
\end{figure*}

\indent To visualize how the shear anisotropy is affected by multiple deflections, we simulate a lens galaxy L1 at a redshift 0.4. We assume the L1 is an SIE, with an ellipticity $e_1=0.2$ and $e_2=0$. The dark matter halo is perfectly aligned with the light distribution, and has the same ellipticity as the lens. We compute the lensing signal with Equation (\ref{eq_sis}) for a velocity dispersion of 200 km s$^{-1}$. In the absence of multiple deflection, we find that $f_{\rm{mm}}=(\pi+e_h)/(\pi-e_h)=1.136$. We insert a second lens L2 in the image, a round SIS at $z=0.1$ with a velocity dispersion $\sigma=200$ km s$^{-1}$. For every position of L2 in the simulated image, we calculate $\tilde{f}_{\rm{mm}}(r=100)$, the shear ratio we would observe for a projected lens-source separation of 100 arcsecs, using a source redshift of 0.8. The result is shown in Figure \ref{plot_multdefl}. \\
\indent We find that depending on the location of L2, $\tilde{f}_{\rm{mm}}$ either becomes larger or smaller than 1.136, and even in some configurations turns negative. These trends can easily be understood: e.g. if L2 is located at (50,250), left to the lens and the ring of sources, it increases the tangential shear of the sources along the major axis (region {\tt B} in Figure \ref{plot_lensschematic}), but it decreases the tangential shear of the sources along the minor axis (region {\tt A} in Figure \ref{plot_lensschematic}). The impact on the ellipticity of L1 is very small, and consequently we find that $\tilde{f}_{\rm{mm}}$ increases. If L2 is located very close to the sources, it can change the source ellipticities by such amounts that the net tangential shear with respect to L1 becomes negative, which results in a negative shear ratio.\\
\indent Equivalently to Equation (\ref{eq_fmmtilde}), we compute $\tilde{f}_{\rm{mm}}^{\rm{corr}}$ as a function of the position of L2, which is shown in the right-hand panel of Figure \ref{plot_multdefl}. If L2 is at a large distance from the sources and L1, the induced shear on both L1 and the sources is almost constant, and is effectively removed using the cross terms. For smaller separations, correcting the ratio using the cross terms does not work well as the applied shear from L2 varies strongly along the ring of sources. \\
\indent To quantify the net effect of multiple deflections on the shear ratio, we determine the total contribution to $\gamma_{t,A}$ and $\gamma_{t,B}$ by integrating over all L2 positions. Then we compute the average value of the ratio:
\begin{equation}
  \langle\tilde{f}_{\rm{mm}}(r)\rangle=\frac{\gamma_{t,B0}+\bar{n_{L2}}\Delta \gamma_{t,B}}{\gamma_{t,A0} +\bar{n_{L2}}\Delta \gamma_{t,A}},
\end{equation}
where $\gamma_{t,A0}$ and $\gamma_{t,B0}$ are the average tangential shear along the minor and major axis in the absence of L2, $\Delta \gamma_{t
,A}$ and $\Delta \gamma_{t,B}$ are the total contributions to the tangential shear along the minor and major axis, and $\bar{n_{L2}}$ is the foreground galaxy number density. The change of the shear ratio depends on the ellipticity of the lens, the density profiles of L1 and L2, the area within which we integrate the contributions of L2, and on the number of second deflectors, $\bar{n_{L2}}$. We study these dependencies below. \\
\begin{figure*}
  \begin{minipage}[t]{0.32\linewidth}
      \includegraphics[width=1.1\linewidth]{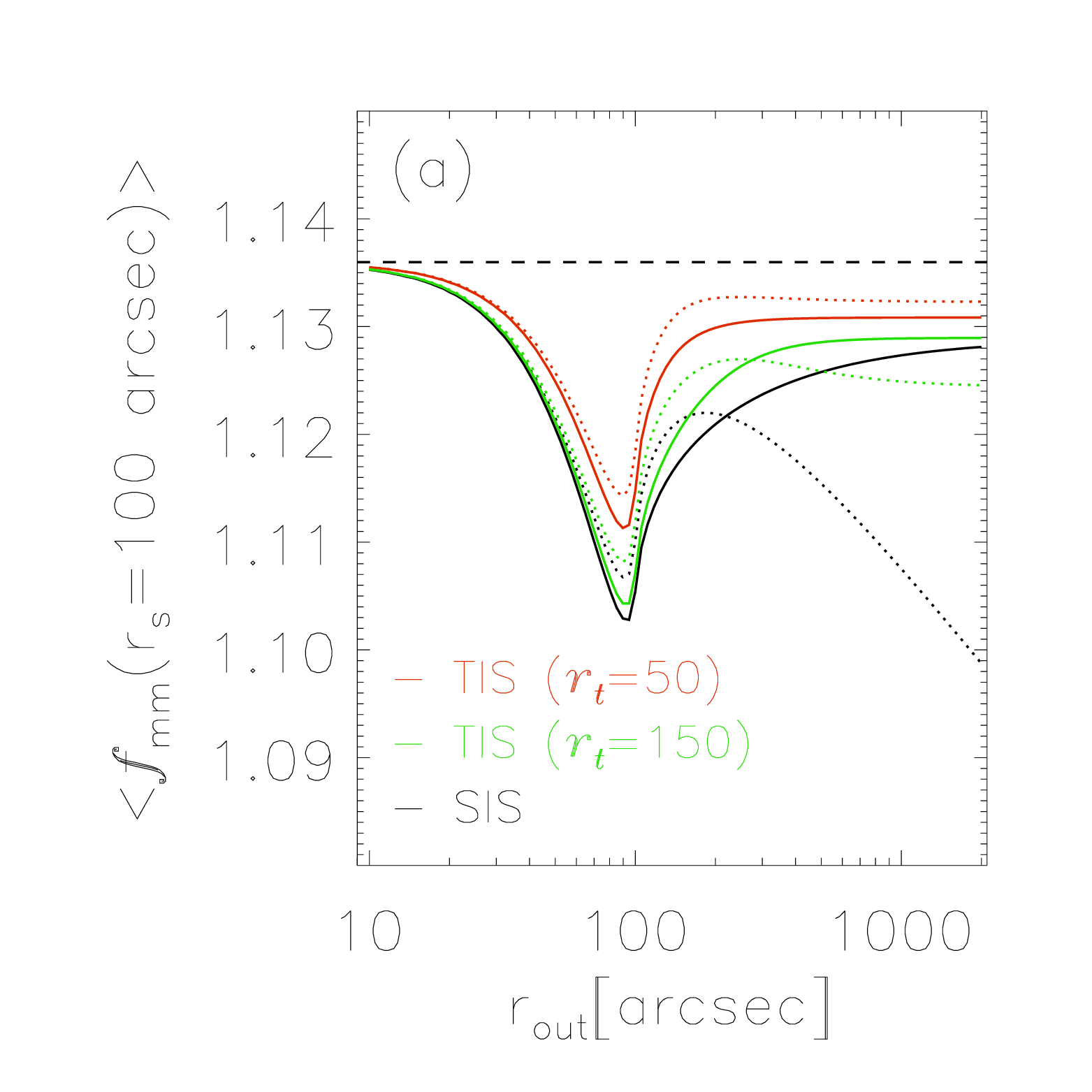}
  \end{minipage}
  \begin{minipage}[t]{0.32\linewidth}
      \includegraphics[width=1.1\linewidth]{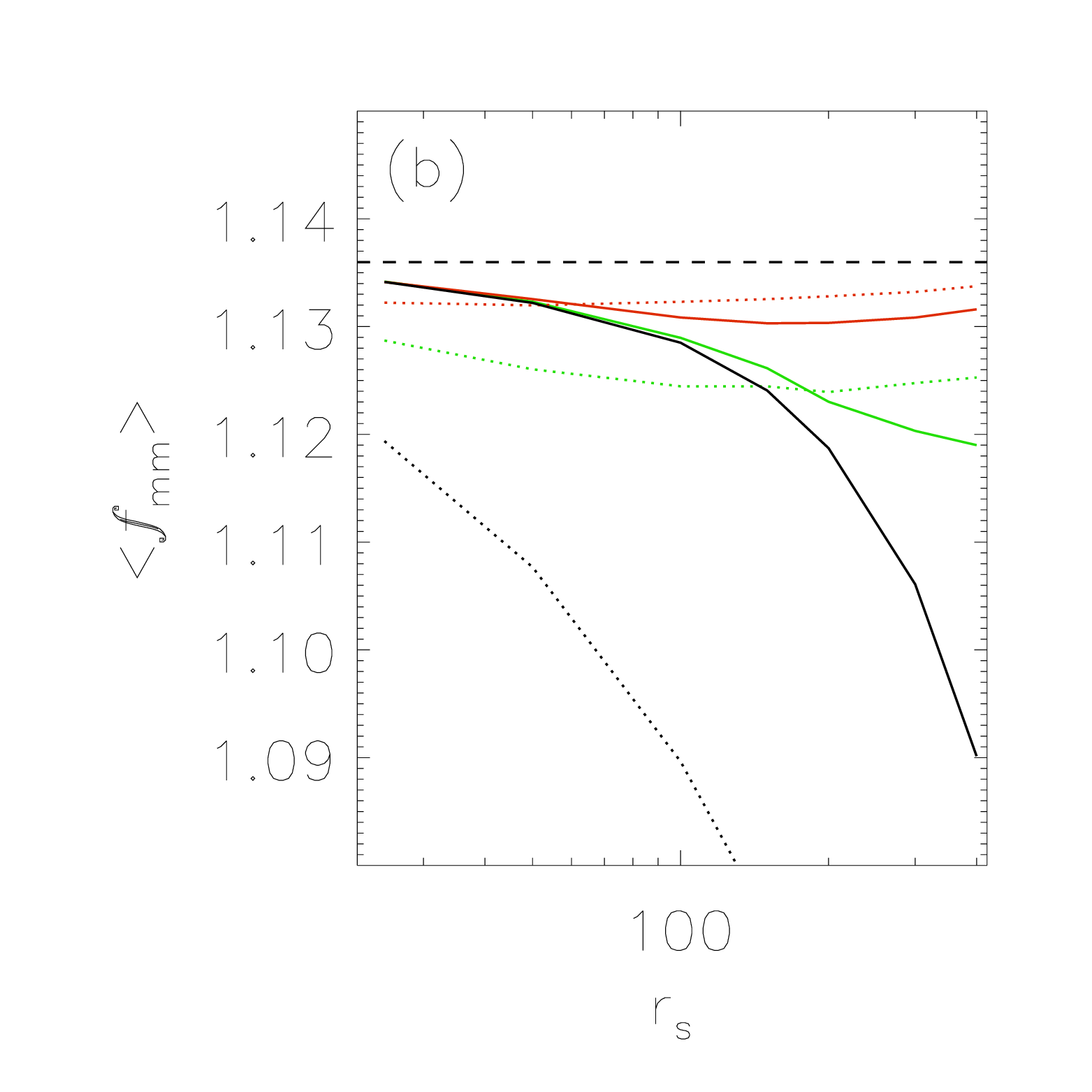}
  \end{minipage}
  \begin{minipage}[t]{0.32\linewidth}
      \includegraphics[width=1.1\linewidth]{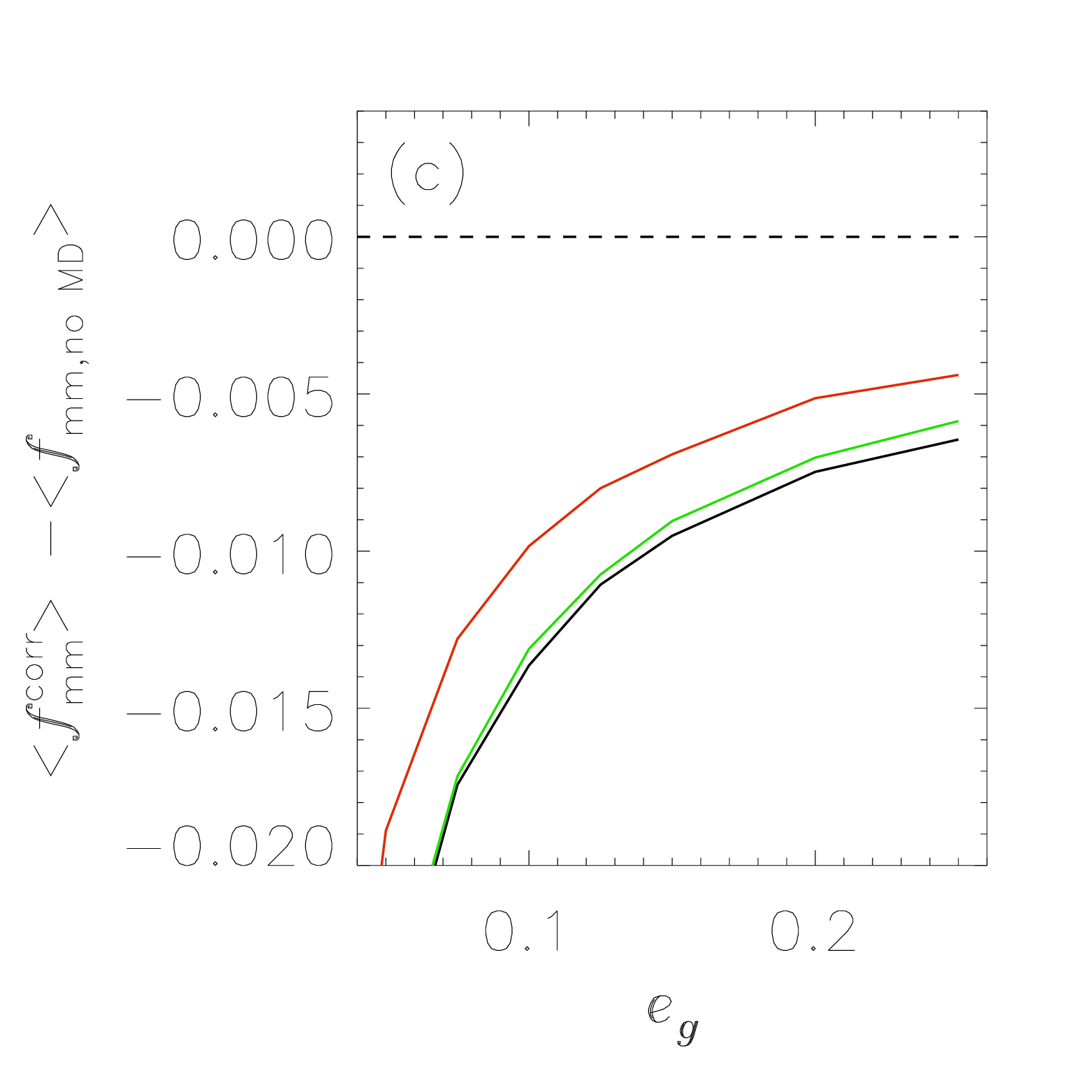}
  \end{minipage}
  \caption{$(a)$ Observed value of ${f}_{\rm{mm}}$ (dotted line) and ${f}_{\rm{mm}}^{\rm{corr}}$ (solid line) for a lens L1 with an SIE profile and ellipticity $(e_1,e_2)=(0.2,0.0)$, measured using sources at 100 arcsec, as a function of $r_{\rm{out}}$, the projected radius from L1 within which the contributions of L2 are integrated. The black lines show the reduction if L2 is an SIS, the green (red) if L2 is a TIS with a truncation radius of 150 (50) arcsec. The dashed line indicates the shear ratio in the absence of L2. This figure demonstrates that L2 at projected separations $>100$ arcsec from L1 have a non-negligible influence on the shear ratio. Multiple deflections reduce the observed shear ratio, but using the cross shear we can correct for most of it. $(b)$ Observed value of ${f}_{\rm{mm}}^{\rm{corr}}$ as a function of source radius, for an $r_{\rm{out}}$ of 4000 arcsec. The reduction of ${f}_{\rm{mm}}^{\rm{corr}}$ increases with increasing L1-source separation. $(c)$ Difference between the observed value of ${f}_{\rm{mm}}^{\rm{corr}}$ and the shear ratio in the absence of multiple deflections ($f_{\rm{mm,no \hspace{1mm} MD}}$) as a function of the ellipticity of L1, measured at a L1-source separation of 100 arcsec, for an outer radius of integration of 4000 arcsec. The reduction of the shear ratio increases with decreasing ellipticity. }
  \label{plot_radchange}
\end{figure*}
\indent In Figure \ref{plot_radchange}a, we show $\langle\tilde{f}_{\rm{mm}}\rangle$ and $\langle\tilde{f}_{\rm{mm}}^{\rm{corr}}\rangle$ as a function of $r_{\rm{out}}$, the radius of the circle centred at L1 within which we integrate the contributions of L2 (hence for a $r_{\rm{out}}$ of 50 arcsecs, we only account for contributions of L2 that are located within 50 arcsecs from L1), for a L1-source separation of 100 arcsec. We use three different profiles for L2, i.e. an SIS and two truncated SIS (TIS) profiles with a truncation radius of 50 and 150 arcsec. Most real galaxies have a density distribution that falls somewhere in between these extremes. The value for $\bar{n_{L2}}$ we have adopted is 100 galaxies per square degree. The velocity dispersions and redshifts of the galaxies are similar to the values used before. \\
\indent We find that the reduction of the shear ratio is a strong function of $r_{\rm{out}}$. It reaches a minimum when $r_{\rm{out}}$ is equal to the L1-source projected separation, which is expected from Figure \ref{plot_multdefl}; if L2 is located at the ring of sources, the shear ratio becomes large and negative, which leads to a reduction of the average ratio. For larger $r_{\rm{out}}$, we find that the shear ratio increases again. If L2 is an SIS, we find that the ratio turns over and continues to decrease. If L2 is a TIS, it does not contribute to the shear ratio if it is located at a projected separation much larger than the truncation radius, and the ratio therefore converges to a certain value. We observe that the impact of multiple deflections is mostly removed for $\langle\tilde{f}_{\rm{mm}}^{\rm{corr}} \rangle$. Even if L2 is an SIS, the total reduction of the shear ratio is small as long as we integrate the contributions of L2 over a sufficiently large area. \\
\indent In Figure \ref{plot_radchange}b, we show $\langle\tilde{f}_{\rm{mm}}\rangle$ and $\langle\tilde{f}_{\rm{mm}}^{\rm{corr}}\rangle$ as a function of projected source separation for an $r_{\rm{out}}$ of 4000 arcsec. We find that $\langle\tilde{f}_{\rm{mm}}\rangle$ decreases with $r_s$ if L2 is an SIS. If L2 is a TIS, $\langle\tilde{f}_{\rm{mm}}\rangle$ reaches a minimum because L2 cannot shear both L1 and the sources if $r_s$ becomes much larger than the truncation radius, which reduces the impact on $\langle\tilde{f}_{\rm{mm}}\rangle$. Furthermore, we find that $\langle\tilde{f}_{\rm{mm}}^{\rm{corr}} \rangle$ is hardly affected by multiple deflections on small scales. On large scales, however, the contribution from L2 is no longer constant, and the correction scheme fails.\\
\indent Finally, we show the bias of the shear ratio as a function of ellipticity of L1 in Figure \ref{plot_radchange}c. We find that the reduction is largest for the roundest lenses, as their position angles are affected most by the presence of L2. The impact of multiple deflections decreases for lenses with larger ellipticities. \\
\begin{figure*}
  \resizebox{\hsize}{!}{\includegraphics[angle=270]{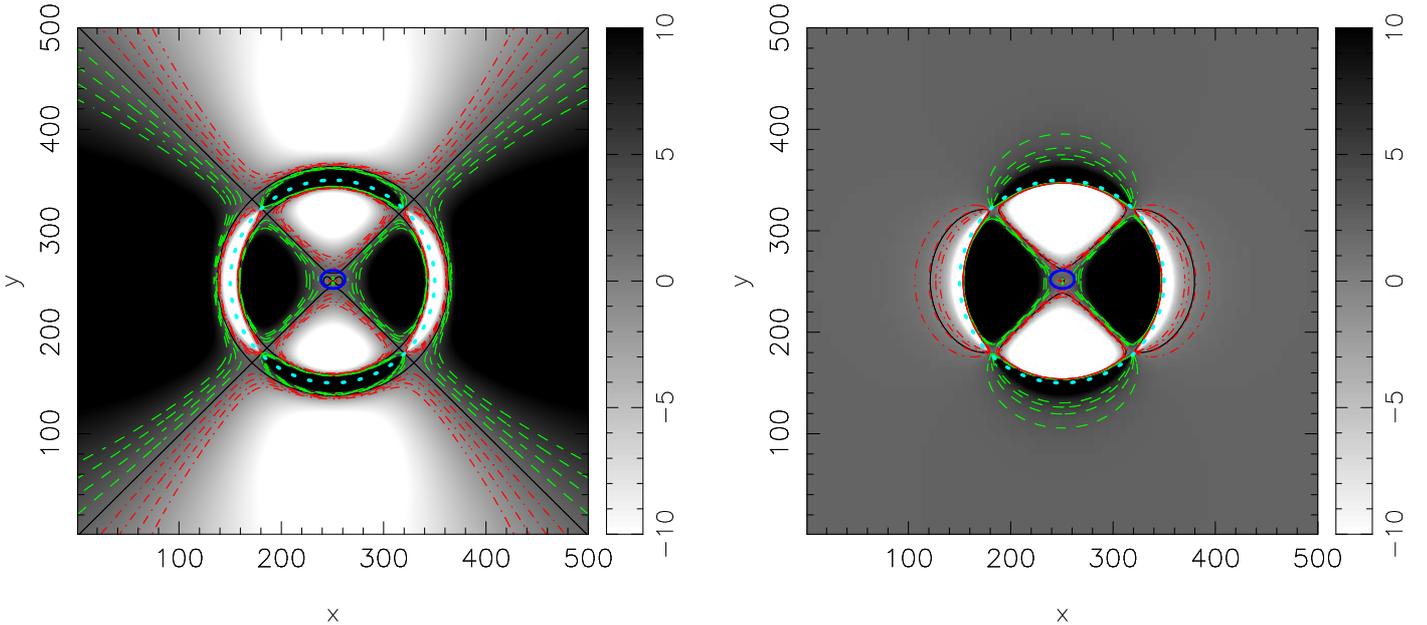}}
  \caption{Net effect on $f \Delta \Sigma_{\rm{iso}}$ (left-hand panel) and $(f-f_{45}) \Delta \Sigma_{\rm{iso}}$ (right-hand panel) for a L1 SIS with an ellipticity $(e_1,e_2)=(0.2,0.0)$, measured using sources at 100 arcsec, in the presence of a second foreground galaxy L2 located at ($x,y$). The location of L1 is indicated by the blue ellipse in the centre of the image, the location of the sources by the cyan dots. We find that depending on the position of L2, the contribution to the shear is either positive or negative. The black solid lines indicate the position of L2 with no net contributions to $f \Delta \Sigma_{\rm{iso}}$ and $(f-f_{45}) \Delta \Sigma_{\rm{iso}}$, the red dot-dashed (green dashed) lines indicate a decrease (increase) of $\Delta \Sigma$ of 1, 2 and 3 [$M_{\odot}$/pc$^2$] respectively.}
  \label{plot_multdefl2}
\end{figure*}
\indent For completeness, we show the impact of an additional foreground lens L2 located at $(x,y)$ on $f \Delta \Sigma_{\rm{iso}}$ and $(f-f_{45}) \Delta \Sigma_{\rm{iso}}$ in Figure \ref{plot_multdefl2}. Similarly as for the shear ratios shown in Figure \ref{plot_multdefl}, we find that the cross terms remove most of the systematic contributions as long as L2 is not located very nearby.

\section{Intrinsic alignments \label{ap_ia}}
The shear anisotropy $(f-f_{45})\Delta\Sigma_{\rm{iso}}$ of the `red' lens sample on small scales is negative (Figure \ref{plot_shearani}). Since the source sample contamination of physically associated galaxies is largest at the same scales (see Figure \ref{plot_od}), and since the distribution of these satellite galaxies is anisotropic (see Figure \ref{plot_odani}), a negative shear signal could also be caused if these satellite galaxies are preferentially radially aligned. To estimate the value of the average tangential intrinsic alignment that would produce such a signal, we rewrite Equation (\ref{eq_falpha}):
\begin{equation}\begin{split}
  f \Delta \Sigma_{\rm{iso}} (r) = \frac{A}{N}\sum_{i=1}^{N_s} w_i \Delta \Sigma_i e^{\alpha}_{g,i} \cos(2\Delta \theta_i) + \\
  \frac{A}{N}\sum_{i=1}^{N_I} w_i \Delta \Sigma_{\rm{crit}} \gamma_I e^{\alpha}_{g,i} \cos(2\Delta \theta_i); 
  \label{eq_fint}
\end{split}\end{equation}
$$ N=2 \sum_{i=1}^{N_s} w_i e_{g,i}^{2\alpha} \cos^2(2\Delta \theta_i)+ 2 \sum_{i=1}^{N_I} w_i e_{g,i}^{2\alpha} \cos^2(2\Delta \theta_i), $$
with $\gamma_I$ the tangential intrinsic alignment, and $N_s$ and $N_I$ the number of sources and physically associated galaxies, respectively. We assume that the shear signal is isotropic, hence the first term on the right-hand side of Equation (\ref{eq_fint}) cancels. As the shear anisotropy is expected to be positive for the red early-type galaxies that make up the `red' lens sample, our estimate of $\gamma_I$ is a lower limit. To account for the fraction of satellites that is anisotropically distributed, we multiply Equation (\ref{eq_fint}) with $f_{\rm{cg}}(r,\Delta\theta)/\langle f_{\rm{cg}}(r)\rangle_{\Delta\Theta}$. For $f_{\rm{cg}}$ we adopt the same form as in Section \ref{sec_contam}, i.e. Equation (\ref{eq_fcg}), which we insert into Equation (\ref{eq_fint}). Ignoring the isotropic part, which cancels when averaged over the angle, we obtain:
\begin{equation}
f \Delta \Sigma_{\rm{iso}} (r) =\frac{A}{N}\sum_{i=1}^{N_I} w_i \Delta \Sigma_{\rm{crit}} \gamma_I e^{2\alpha}_{g,i} \cos^2(2\Delta \theta_i)\frac{2N_{\Delta\theta}}{N_{\rm{iso}}-1}, \\
\end{equation}
and therefore
\begin{equation}\begin{split}
\gamma_I=\frac{f \Delta \Sigma_{\rm{iso}} (r)}{2A\Sigma_{\rm{crit}}}\frac{N_s+N_I}{N_I}\frac{N_{\rm{iso}}-1}{N_{\Delta\theta}} \\
=\frac{f \Delta \Sigma_{\rm{iso}} (r)}{2A\Sigma_{\rm{crit}}}\frac{N_{\rm{iso}}}{N_{\Delta\theta}},
\end{split}\end{equation}
where we used that $(N_s+N_I)/N_I = N_{\rm{iso}}/(N_{\rm{iso}}-1)$. Averaging the results of the first two radial bins of the `red' lenses, we obtain $\gamma_I=-0.062_{-0.062}^{+0.042}$ for $\alpha=1$ at an average separation of 0.19 arcmin ($\sim$65 kpc at the mean lens redshift). Our constraints are not particularly competitive (compare, e.g., Hirata et al. 2004), mainly because our methods are not designed for this measurement. For example, had we included brighter galaxies in the source sample, a larger fraction of those would be physically associated to the lenses, improving the constraints on the galaxy overdensity, and hence on $\gamma_I$. This, however, is not the purpose of this work. Note that the value and error on $\gamma_I$ we obtain is roughly a factor 10 larger than the results from \citet{Hirata04}, who aimed their analysis to measure this effect using data from the SDSS (although no division was made between early-type and late-type galaxies, which could average out the effect). Conversely, the expected impact of intrinsic alignments is about a factor ten smaller than the signal we observe, hence it is unlikely that our measurements are significantly affected. Hence the negative shear anisotropy is unlikely caused by intrinsic alignments. \\
\end{appendix}


\end{document}